\documentclass[letterpaper,twocolumn,10pt]{article}
\usepackage{usenix}

\usepackage{authblk}
\usepackage{booktabs}
\usepackage{comment}
\usepackage{graphicx}
\usepackage{amsmath,amssymb,amsfonts}
\usepackage{subfig}

\usepackage{enumitem}
\usepackage{multirow}
\usepackage{float}
\usepackage{xspace}
\usepackage{adjustbox}
\usepackage{listings}
\usepackage{appendix}
\usepackage{setspace}

\newcommand{\eg}{e.g.,\xspace}
\newcommand{\ie}{i.e.,\xspace}

\usepackage{algorithmicx,algorithm}
\usepackage[noend]{algpseudocode}
\algdef{SE}[DOWHILE]{Do}{doWhile}{\algorithmicdo}[1]{\algorithmicwhile\ #1}%

\newcommand{\gl}[1]{\textcolor{red}{#1}}
\newcommand{\sys}{{\textit{EmTaint}}\xspace}

\renewcommand{\paragraph}[1]{\vspace{5pt}\noindent\textbf{#1}}

\newcommand{\cexpr}{{$CTexpr$}\xspace}
\newcommand{\pexpr}{{$Pexpr$}\xspace}

\definecolor{codegreen}{rgb}{0,0.6,0}
\definecolor{codegray}{rgb}{0.5,0.5,0.5}
\definecolor{codepurple}{rgb}{0.58,0,0.82}
\definecolor{backcolour}{rgb}{0.95,0.95,0.92}
 
\lstdefinestyle{mystyle}{
    commentstyle=\color{codegreen},
    keywordstyle=\color{magenta},
    numberstyle=\tiny\color{codegray},
    stringstyle=\color{codepurple},
    basicstyle=\footnotesize,
    breakatwhitespace=false,         
    breaklines=true,                 
    captionpos=b,                    
    keepspaces=true,                 
    numbers=left,                    
    numbersep=5pt,                  
    showspaces=false,                
    showstringspaces=false,
    showtabs=false,                  
    tabsize=2
}
\begin{document}






\title{ \Large \bf Finding Taint-Style Vulnerabilities in Linux-based Embedded Firmware with SSE-based Alias Analysis}

\author[1,2]{\rm Kai Cheng}
\author[3]{\rm Tao Liu}
\author[4]{\rm Le Guan}
\author[3]{\rm Peng Liu}
\author[1]{\rm Hong Li}
\author[1]{\rm Hongsong Zhu}
\author[1]{\rm Limin Sun}

\affil[1]{\small \textit {Institute of Information Engineering, Chinese Academy of Sciences, China}}
\affil[2]{\small \textit {School of Cyber Security, University of Chinese Academy of Sciences, China}}
\affil[3]{\small \textit {The Pennsylvania State University}}
\affil[4]{\small \textit{University of Georgia} 
\authorcr 
{chengkai@iie.ac.cn, tul459@psu.edu, leguan@uga.edu, pxl20@psu.edu, \{lihong, zhuhongsong, sunlimin\}@iie.ac.cn}}



\maketitle



\begin{abstract} 
Although the importance of 
using static analysis to detect taint-style vulnerabilities
in Linux-based embedded firmware is widely recognized, existing approaches are plagued by three major limitations. 
(a) Approaches based on symbolic 
execution may miss alias information and therefore suffer from a high false-negative rate.
(b) Approaches based on VSA (value set analysis) often provide an over-approximate pointer range.
As a result, many false positives could be produced. (c) Existing work for detecting taint-style vulnerability 
does not consider indirect call resolution, whereas indirect calls
are frequently used in Internet-facing embedded devices. As a result, many false negatives could be produced.

In this work, we propose a precise demand-driven flow-, context- and field-sensitive alias analysis approach.
Based on this new approach, we also design a novel indirect call resolution scheme.
Combined with sanitization rule checking, our solution 
discovers taint-style vulnerabilities by static taint analysis.
We implemented our idea with a prototype called \sys and
evaluated it against 35 real-world embedded firmware samples from six popular vendors.
\sys discovered at least 192 bugs, including 41 n-day bugs and 151 0-day bugs.
At least 115 CVE/PSV numbers have been allocated from a subset of the reported vulnerabilities at the time of writing.
Compared to state-of-the-art tools such as KARONTE and SaTC,
\sys found significantly more bugs on the same dataset in less time.
\vspace{-2.0mm}
\end{abstract}

\section{Introduction}
\label{intro}
\vspace{-2.0mm}
With the emerging of the \textit{Internet of Things} (IoT) 
technologies, Linux-based embedded firmware is 
nowadays playing an increasingly important role.  
For example, almost all the mainstream wireless routers 
are running Linux-based embedded firmware under the hood. 
Since wireless routers are often the perimeter defense
that separates the information processing and storing requirements
(e.g., laptops, smartphones, database servers) in 
an apartment, a house or a small business office from 
the Internet, the embedded firmware holds a 
privileged position in home and small business 
networks and therefore, they must operate securely. 
The addition of various IoT devices (e.g., 
thermostat, smart light, smart lock, smart alarm system)
to home and small business networks makes the position held by 
Linux-based embedded firmware even more vital. 
Unfortunately, Linux-based embedded firmware still suffers from 
a number of vulnerabilities in the real world. 

To test the real-world Linux-based embedded firmware,
static analysis~\cite{shoshitaishvili2015firmalice, gotovchits2018saluki, redini2020karonte, cheng2018dtaint} and 
dynamic analysis~\cite{muench2018you, zheng2019firm, ZaddachBFB14} are two basic approaches. 
Recently, substantial progress has been made on automated dynamic analysis of embedded firmware. For example, 
Firmadyne~\cite{chen2016towards} has successfully 
emulated the execution of over 1,900 firmware images. 
However, existing dynamic analysis techniques are still quite limited. 
For example, due to reasons such as 
complex environment dependencies of the firmware,
Firmadyne~\cite{chen2016towards} was only able
to emulate 23\% of the collected 8,591 firmware images.

Being complementary to dynamic analysis, static analysis 
of embedded firmware has also been attracting an increasing 
interest in the research community. A unique merit
of static analysis is that in many cases it can 
simply ignore the complex environment dependencies of the firmware
while still being able to find many vulnerabilities.  
In this work, we propose a new static binary analysis approach to find 
taint-style vulnerabilities in COTS (commercial off-the-shelf)  Linux-based embedded firmware.
(Because the source code is not available, note that 
  we cannot use any source-code-analyzing tools.)  
The taint-style vulnerability is a class of vulnerabilities  
that share the same underlying theme rooted in 
information flow analysis~\cite{automaticinference}.
More specifically, attacker-controlled data are passed from an input source to 
a security-sensitive sink without proper check or sanitization.
This vulnerability class captures the root cause for many software defects and
is often manifested into many common vulnerability types including
buffer overflow and format string vulnerability. 

Given an embedded firmware image, 
the standard procedure for detecting taint-style vulnerability
consists of four steps.
(1) Recover the inter-procedural control flow graph (ICFG)
of the program.
(2) Identify attacker-controlled sources and security-sensitive
sinks.
(3) Find a path where the taint is propagated from
the source to the sink.
(4) Check the constraints of the tainted data at sinks. If not
constrained, an alert about the vulnerability is raised.
An ideal solution should achieve high accuracy in all the steps.
When high accuracy is not achieved, the analysis usually suffers from 
two bad consequences: (1) a good portion of the paths 
found in Step 3 hold one or more {\bf bogus} links: such paths would produce false positives; 
(2) due to missing links, some vulnerability-revealing paths 
are not identified: this produces false negatives.

\paragraph{Problems with existing work.} 
As we will discuss in detail shortly in Section 2, we found that 
the existing static binary analysis 
works on taint-style vulnerability hunting are
still very limited in achieving high accuracy, mostly due to the challenges they face in the aforementioned Step 1 and Step 3. Their main limitations can be summarized as follows.

(1) With two alias pointers being dereferenced
to different addresses, works (e.g., KARONTE~\cite{redini2020karonte}) based on symbolic 
execution may suffer from high false negative rate.

(2) Works (e.g., Loongchecker~\cite{cheng2011loongchecker}) based on VSA (value set analysis) often provide an over-approximate pointer range.
As a result, many false positives could be produced.

(3) Existing works for detecting taint-style vulnerability do not consider indirect call resolution, partially due to the associated difficulties.
However, based on our study, indirect calls are frequently
used in Internet-facing embedded devices (see Section~\ref{section:icall}).
By missing many taint propagation links, many false negatives could be produced. 

\vspace{-1.0mm}
\paragraph{Our solution.} We argue that inadequate alias analysis and
lack of indirect call resolution are two technical issues for  cost-effective 
embedded firmware analysis. Crossing these barriers will unleash
the capabilities of static analysis to find taint-style vulnerabilities.
Note that the results of alias analysis can be directly used to recover
indirect calls.
Therefore, in this work, we propose a precise demand-driven flow-, context- and
field-sensitive alias analysis mechanism.
We further leverage it to
find the alias relationship between the indirect call
targets and address-taken functions, yielding an accurate set of
indirect-call targets. 
This is fundamentally different from existing type
matching based approaches which suffer from high false-positives. Combined
with sanitization rule checking, our solution finds more bugs in less time
with fewer false positives.

The proposed SSE-based alias analysis overcomes the drawbacks of existing
VSA-based or symbolic-execution-based approaches in two aspects. First, we
extend the symbolic values used in symbolic execution to incorporate abstract
memory operations, eliminating the need for the information-losing
concretization process. Second, our analysis is demand-driven in a sense that
only interesting variables need to be traced, avoiding the holistic analysis
used in VSA (while localized VSA is possible, it may miss many useful
information and lead to bogus links in general). Concretely, given a variable
at a particular program point, our approach finds all its aliases along the
path by following the use-define and define-use information forwards and
backwards. To execute this design, we borrow ideas from access path~\cite{cheng2000modular}
which represents memory locations by how they are accessed
from an initial variable. Since access path is designed for analyzing source
code, we specifically accommodate it for binary-only analyses by
incorporating our new contributions. Specifically, we propose a new notation
called \textit{structured symbolic expression} (SSE) to facilitate the idea.
SSE encodes the provenance of a variable using symbolic expression. However,
different from symbolic values used in symbolic execution, it not only
encodes arithmetic operations, but also considers memory operations such as
load and store. As such, multi-level pointer dereferences (e.g., {\tt x.y.z})
can be expressed in an SSE as a concatenation of symbolic
load/store/arithmetic statements with base and offset information. Our
algorithm starts from an interesting SSE-encoded variable. When we find a new
use or define for any field in the SSE, a new alias SSE is generated by
replacing that field. The new alias is then used in a new round of alias
searching until a fixpoint is reached. 
SEE tracking can easily traverse across function boundaries, making
inter-procedural analysis straightforward. As such, our approach is flow-,
context-, and field-sensitive. This unique design brings about the following
desirable benefits. First, alias searching can start at any program location
without losing alias information before it. Therefore, it finds a
comprehensive list of aliases for a given variable, regardless of the
specified starting point. In this way, we do not need to rely on expert input
to start analysis from a firmware-specific point near the sink~\cite{cheng2018dtaint,redini2020karonte}. Rather, we can start from commonly used
taint sources such as the {\tt recv} function. Second, our approach does not
need to analyze the program as a whole like VSA. It focuses on a particular
variable and only computes over relevant instructions. This on-demand feature
allows for more efficient and scalable analysis for complex programs.

We have implemented the proposed SSE-based demand-driven alias analysis  and
indirect call resolution based on \textsl{Claripy}. They were integrated into a
prototype system called \textbf{\sys} to find taint-style vulnerabilities in
Linux-based embedded firmware.
We evaluated \sys with 35 real-world firmware samples.
The result shows that \sys quickly produces a large number of alerts.
Regarding indirect call resolution, \sys recovered 12,022 of 12,446 (96.6\%) indirect calls
in an hour.
Regarding vulnerability discovery,
\sys discovered at least 192 bugs, including 41 n-day bugs and 151 0-day bugs.
Each sample takes an average of 3 minutes.
We have reported 151 0-day bugs to the relevant manufacturers for
responsible disclosure. 
115 of which are confirmed by CVE/PSV at the time of writing.
We also conducted a comparison with KARONTE~\cite{redini2020karonte}
and SaTC~\cite{chen2021sharing}.
The results of the comparison shown that \sys can find more bugs in less time.



\vspace{-1mm}
\paragraph{Contributions.} In summary, we make the following contributions in this paper:
\vspace{-2mm}
\begin{itemize}[leftmargin=*]
\item We propose a new demand-driven alias analysis technique based on structured symbolic expressions. With structured symbolic  expressions,
our approach achieves flow-, context- and field-sensitive alias analysis simultaneously. 

\vspace{-2mm}
\item We observe that resolving indirect calls is critical for finding
taint-style vulnerabilities in embedded firmware. Therefore, we propose a novel
indirect call resolution scheme. Our approach directly finds  the alias
relationship between the indirect call targets and address-taken functions with
high confidence.

\vspace{-2mm}
\item We have implemented the proposed system and evaluated it with extensive
experiments. 
Our tool have identified 151 0-day vulnerabilities in 35
embedded firmware samples. At least 115 CVEs/PSVs have been allocated from a subset of 
the reported bugs.
For continuing research, we will open source our tool.
\vspace{-3.0mm}
\end{itemize}








\begin{figure}[t]
\centering 
    \includegraphics[width=\columnwidth]{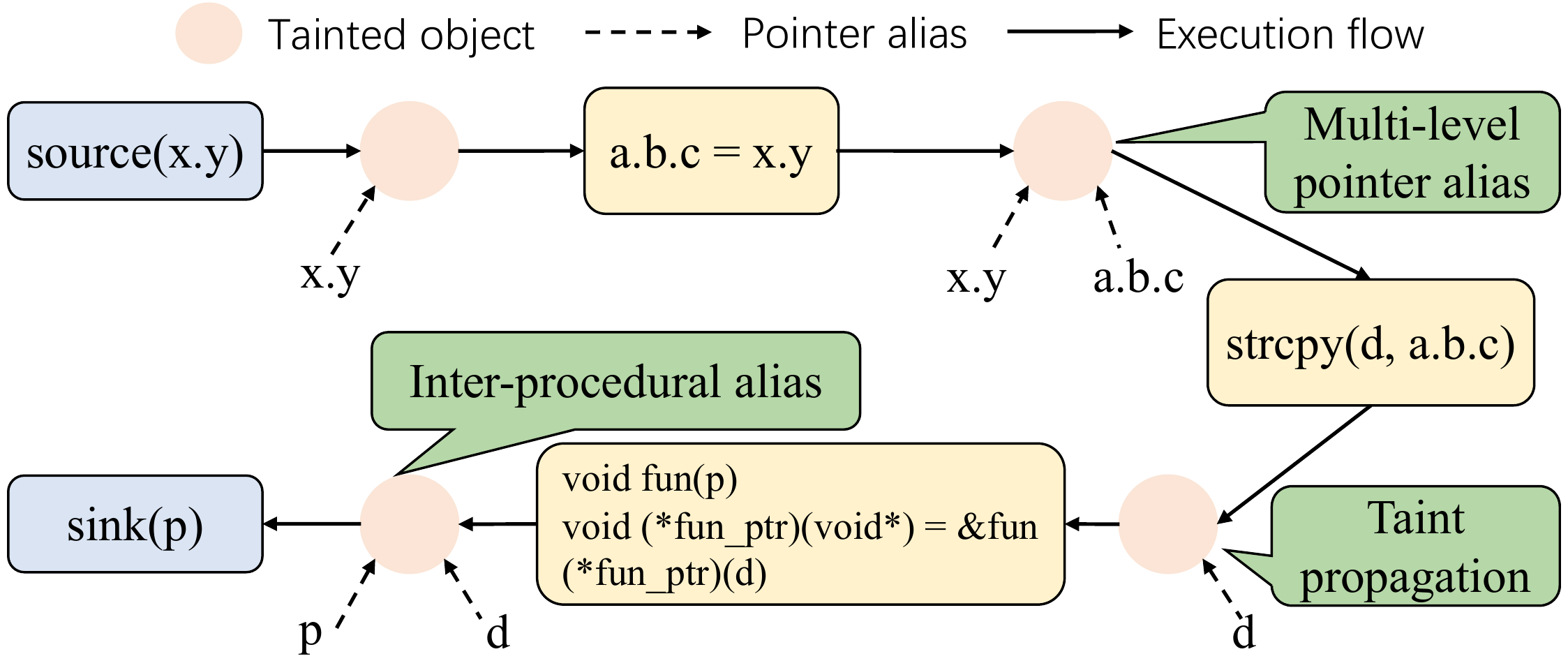}
    \setlength{\abovecaptionskip}{-2.0mm}
    \caption{Taint-style vulnerabilities and how they can be identified in general.}
    \label{fig:taint-style}
    \vspace{-6.0mm}
\end{figure}

\vspace{-3mm}
\section{Background and Motivation}
\label{section:two}

\vspace{-2.0mm}
\subsection{Taint-style Vulnerability}
\vspace{-1.0mm}
The taint-style vulnerability is a class of vulnerabilities  
in which 
attacker-controlled  data  are  passed  from  an  input  source  to
a  security-sensitive  sink  without  sanitization.
Figure~\ref{fig:taint-style} illustrates the data flow of a representative
taint-style vulnerability.
Attacker-controlled data enter the program via the input function
\texttt{source()} and the input buffer is pointed to by \texttt{x.y}.
Since this buffer is controlled by the attacker,
we mark it as tainted.
Then, through an assignment statement, the tainted pointer
makes an alias \texttt{a.b.c}, which is also a multi-level
pointer.
Next, a \texttt{strcpy} function is used to copy the data pointed to by \texttt{a.b.c} to another buffer pointed to by \texttt{d}.
Obviously, the pointer \texttt{d} should be tainted because
it points to the memory containting the same data as the taint source.
Note that \texttt{strcpy} is not only
used to propagate a taint but also a notable characteristic of the vulnerability.
An indirect call is then invoked with \texttt{d} as an argument.
Inside the indirect call, \texttt{d} makes an alias \texttt{p}.
Finally, \texttt{p} is used in a sensitive sink function.

As shown above, the data-flow of a real-world taint-style vulnerability
is quite complex.
The taint could be propagated through \textbf{\emph{multi-level pointer aliases}} and data movement
functions. Sometimes the propagation can even 
traverse through \textbf{\emph{functions that are indirectly invoked}}.
Given an embedded program, the standard procedure for detecting taint-style
vulnerability consists of four steps.
We list each step and discuss the key requirements for them as follows.


1) Recover the \textit{inter-procedural control flow graph} (ICFG)
of the program.
ICFG is the union of all functions' CFG by connecting calling edges.
Compared with direct calls, recovering indirect calls
is considered challenging in static analysis since the calling
target is only determined at run-time.
Recovering the ICFG is necessary for inter-procedural alias analysis.

2) Identify attacker-controlled sources and security-sensitive sinks.
Both the source and sink are typically recognized based
on semantics of the relevant functions.
For example, the function \textit{recv} receives data from the network
and its second buffer pointer should be tainted.
The function \textit{system} invokes a shell to execute
arbitrary commands.
Therefore, its parameter is considered as a sink.
This step requires the accurate identification of the corresponding functions.

3) In the recovered ICFG, find a path where the taint is propagated from the source to the sink. 
This step is the most crucial but challenging. 
As shown in Figure~\ref{fig:taint-style},
it should cross three barriers --
multi-level pointer alias analysis, inter-procedural alias analysis,
and taint propagation analysis.

4) Check the constraints of the tainted data at sinks. 
If not constrained,
an alert about the vulnerability is raised.



An ideal solution should achieve high accuracy in all the steps.
\emph{When high accuracy is not achieved, the analysis usually suffers from 
two bad consequences: (1) a good portion of the paths 
found in Step 3 hold one or more {\bf bogus} links: such paths do not reveal 
real vulnerabilities, they produce false positives; 
(2) due to missing links, some vulnerability-revealing paths 
are not identified: this produces false negatives. }





\vspace{-1.2em}
\subsection{Limitations of Existing Techniques}
\vspace{-2.0mm}
Unfortunately, existing works on taint-style vulnerability hunting are
still very limited in achieving high accuracy, mostly due to the challenges they face in  
the aforementioned Step 1 and Step 3.



\vspace{-3.0mm}
\subsubsection{Alias Issues}

\vspace{-3.0mm}

\paragraph{Works based on symbolic execution may suffer from high
false negative rate.} Symbolic execution has been
widely used in taint
analysis~\cite{cova2006static,redini2017bootstomp,redini2020karonte,gotovchits2018saluki}.
Concretely, existing works taint the input and propagate the tainted data via
forward symbolic execution. An alert is raised when a feasible path is found
to have the taint in any of the sinks without sanitization.  Apart from the well-recognized
problems in symbolic execution (path explosion and timing-consuming constraint
solving), the accuracy of these approaches depend on the concretization
strategy. Specifically, when a symbolic address is accessed, the symbolic
execution engine has to concretize the symbol to allow for continuous
execution. Ideally, two alias symbolic addresses should be concretized into
the same concrete address. However, it is a non-trivial task to maintain this
information in symbolic execution. With two alias pointers being dereferenced
to different addresses, taints cannot propagate, causing a high false negative
rate.

To mitigate path explosion issues, existing
works~\cite{redini2017bootstomp,redini2020karonte} use under-constrained
symbolic execution~\cite{ramos2015under}, in which symbolic execution starts
from arbitrary functions, instead of the main entry point. In this way, the
distance from the source to the sink is reduced and thus the chance for
indirect calls along the path is decreased.  However, this approach {\bf needs experts to manually} specify 
firmware specific source functions.

\paragraph{Works based on VSA often provide an over-approximate pointer range.} 
VSA has been widely used in many static
analysis applications, ranging from recovering binary
properties~\cite{balakrishnan2007divine,balakrishnan2005recovery} to
identifying binary vulnerabilities~\cite{rawat2011static,cheng2011loongchecker}. The high-level idea
of VSA is to maintain a tight over-approximation of the set of numeric values
or addresses that each register or variable \emph{might} hold at a given
program point. By providing information about the register values that appear
in an indirect memory operand, VSA can determine the addresses that are
potentially accessed. This is particularly useful for alias analysis which
aims to determine whether two pointers refer to the same
address~\cite{reps2008improved}. However, when VSA
computes the pointer ranges by following a set of rules (e.g., add operation),
it can only provide an over-approximate range. Alternatively, when VSA has no
clue to constrain a pointer, the pointer becomes even untrackable. This
problem becomes noticeable when it comes to complex programs. Specifically,
bogus dependency renders it expensive and unpractical for real-world
programs~\cite{zhang2019bda}.

\vspace{-1em}
\subsubsection{Indirect Call Issues}
\vspace{-1mm}
{\bf Existing works for detecting taint-style vulnerability do not consider
indirect call resolution, partially due to the associated difficulties.}  Most
related works focus on resolving indirect jumps (e.g., switch statements in
C) or simple indirect calls~\cite{meng2016binary, cifuentes2001recovery,
de2000static, reinbacher2011precise}, in which
no cross-function references occurs.
None of them can effectively handle
complex situations such as indirect calls implemented by callbacks or
function tables. For example,  X-Force~\cite{peng2014x} uses concrete
execution to force binary execution without requiring valid inputs or proper
environment. However, it is expensive and introduces many infeasible paths.
TypeArmor~\cite{van2016tough} infers the function type information from binary
and uses argument number to match callers and callees. However, it suffers
from a high false-positive rate because many functions may share the same
type. To refine the indirect-call targets,  TypeDive~\cite{lu2019does}
leverages multi-layer type information. However, this approach requires the
source code.

\vspace{-1.3em}
\section{Overview}
\label{section:3}

\vspace{-3mm}

\subsection{Key Insight}
\label{sec:insight}

Both symbolic execution based and VSA based approaches face challenges to find
accurate alias information for real-world programs. 
The root cause is that both techniques use a store-based approach to represent runtime memory locations,
which cannot accurately track indirect memory accesses (\ie~\texttt{x.y.z}) whose address cannot be unambiguously represented.
For example, when reading from an unknown location, VSA conservatively models the response as any value (\ie~$\top$),
while symbolic execution models the response as an unconstrained symbolic value.
Both introduce inaccuracy when this value is later used as
an address (\eg~symbolic execution must concretize it using certain strategies).
To address this problem, the store-less approach can be used,
in which memory locations (or pointers) are represented
by how they are accessed from an initial variable (\ie~provenance),
instead of a value (\eg~a-Locs in VSA and symbolic or concrete values in symbolic execution).
This idea was initially proposed in access path to find aliases
in source code~\cite{cheng2000modular}. We are inspired by it and incorporate
into it our new contributions to handle binaries.
The key contribution of this work is the 
newly proposed structured symbolic expression (SSE),
which works at instruction (or IR) level and can represent arbitrary computations, including how a pointer accesses a memory location.
Therefore, it does not rely on the source code information
and directly handles multi-level pointers.
Using SSE, we
can derive new aliases of a variable of interests by iteratively traversing
the define-use and use-define chain in a binary. At each define or use point,
a new alias is derived, represented by another SSE that encodes the
corresponding provenance information.

\begin{figure}[t]
\centering
\includegraphics[width=0.70
\columnwidth]{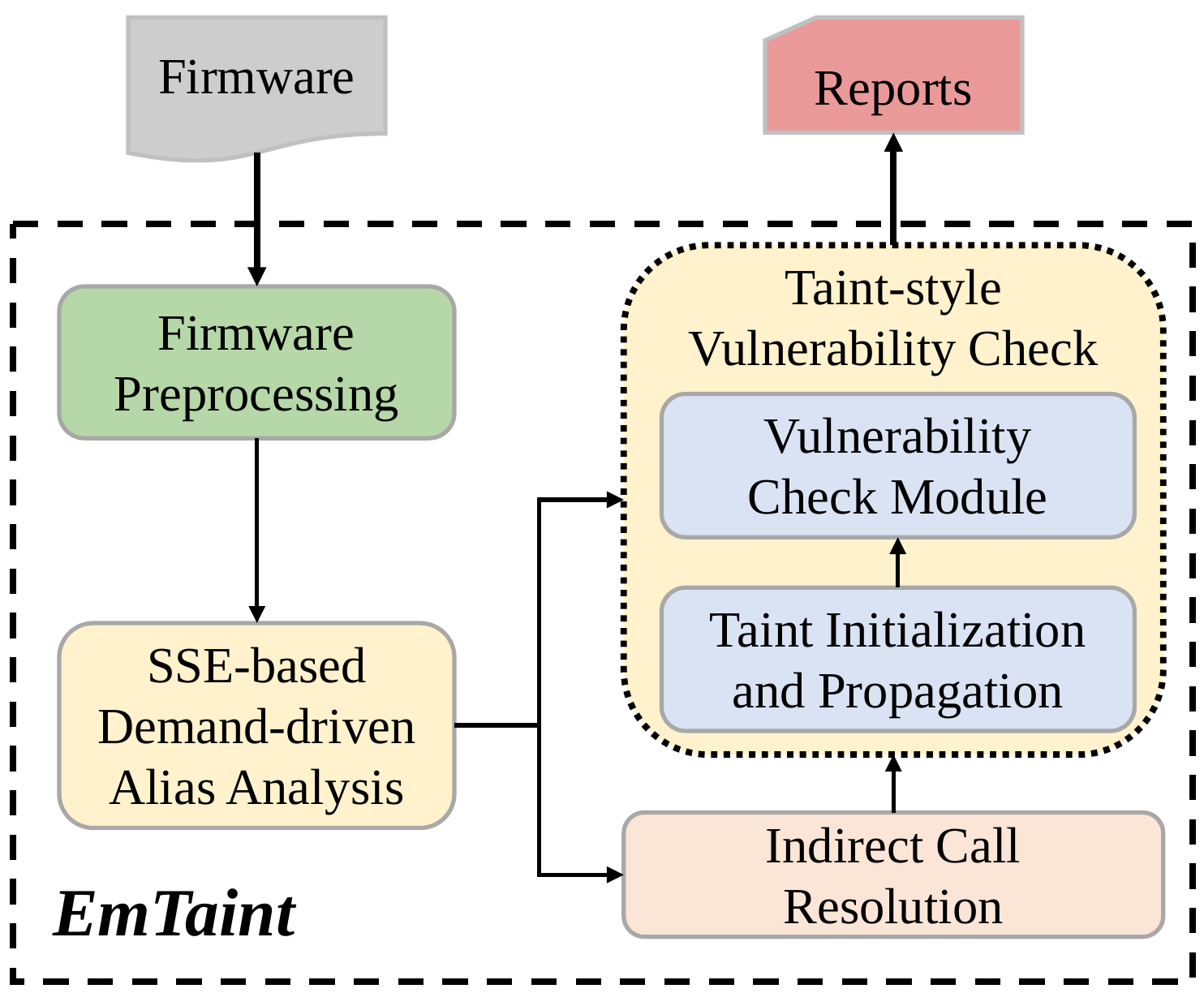}
\setlength{\abovecaptionskip}{1.0mm}
\caption{\sys Overview} 
\label{fig:overview}
\vspace{-7.0mm}
\end{figure}

\vspace{-1em}
\subsection{Architecture Overview}
\vspace{-1mm}
\sys takes an embedded firmware image as input and reports
potential taint-style vulnerabilities through static analysis.
As shown in Figure~\ref{fig:overview}, the proposed system is comprised of four major components.
\vspace{-1mm}

\paragraph{Firmware preprocessing.}
The firmware preprocessing module utilizes \textit{Binwalk}~\cite{binwalk} 
to decompress and extract binaries from firmware
and leverages the state-of-the-art reverse 
engineering tools to extract code and data from the binaries and converts
them into intermediate representation (IR).
It also builds the control flow graph (CFG) and a partial call graph (CG)
that facilitate static analysis (Section~\ref{section:pre}).

\vspace{-1mm}

\paragraph{SSE-based demand-driven alias analysis.}
The SSE-based alias
analysis engine is a novel flow-, context- 
and field-sensitive alias analysis tool (Section~\ref{section:alias}).
It is demand-driven, therefore it scales well for complex programs.
This component lays the foundation for efficient taint tracking 
and indirect call resolution mechanisms.

\vspace{-1mm}

\paragraph{Indirect call resolution.}
Leveraging the data dependence information
provided by the proposed SSE-based alias analysis engine, 
we further design an indirect call resolution algorithm
that checks the data dependence between the referenced function pointers (or function table pointers) 
and target pointers in indirect callsites (see Section~\ref{section:recovery}).
Since indirect call resolution allows tainted variables
to traverse more function boundaries,
it positively impacts the 
comprehensiveness of the data flow analysis in \sys 
and more importantly substantially improves the 
discovery of taint-style vulnerabilities.
Given the pervasiveness of indirect calls in firmware,
the indirect call resolution module is one of the key enablers for
the effectiveness of the proposed solution.
\vspace{-1mm}

\paragraph{Taint-style vulnerability check.}
The taint-style vulnerability check module is responsible for
identifying the potential taint-style vulnerabilities.
Specifically, it firstly taints the variables at taint sources
that are controlled by attackers (e.g., the \textit{recv} function).
Then, it captures taint propagation via the demand-driven alias analysis module
and taint propagation functions (\eg~\texttt{strcpy}).
Finally, at security-sensitive sinks (e.g., the \textit{system} function),
if the corresponding variable is tainted and unconstrained,
we mark the source-to-sink path as potentially exploitable (see Section~\ref{section:taint}). 
\vspace{-1em}
\section{System Design and Implementation}
\label{section:design}
\vspace{-2mm}
In this section, we illustrate the detailed design
and implementation
for each component.
The proposed analysis is based on the VEX
\textit{intermediate representation} (IR)~\cite{nethercote2007valgrind}.
The VEX IR is a popular IR widely used
in many program analysis tools, including
Valgrind~\cite{nethercote2007valgrind} and angr~\cite{shoshitaishvili2016sok}.
It is
architecture-agnostic, so it can be translated from a number of target machine languages, including x86, ARM, MIPS, and PowerPC.
Benefiting from this design choice, our tool
is applicable to mainstream architectures used in embedded firmware,
including ARM and MIPS. 

\vspace{-1em}
\subsection{Firmware Preprocessing}
\label{section:pre}
\vspace{-2mm}
An embedded firmware image is typically a Binary Large OBject (BLOB),
which is a collection of binary data that includes both
the (compressed) kernel image and the file system.
First, we use \textit{Binwalk}~\cite{binwalk} to 
recognize, unpack, and extract executables of our interest (e.g.,~\texttt{httpd}). 
Then, we use IDA Pro~\cite{empiricalstudy} to automatically identify code/functions for
the selected executables.
However, IDA Pro is not proficient enough to locate
some functions that are indirectly called.
We therefore developed an IDA plugin to augment this capability.
Our observation is that many
function pointers are hardcoded in the data segment. 
Therefore, our plugin 
scans the data segment and collects all the immediate values as candidates.
To confirm a function pointer, a candidate must meet two conditions.
One is that its value has to fall into a code segment, and the other is that
the instruction sequence where the candidate points to
must match the signature of a function prologue. 
After obtaining all the potential functions,
\sys loads the binary with APIs provided by angr~\cite{shoshitaishvili2016sok}, 
converts them into VEX
IR by using \textit{pyvex}~\cite{pyvex} and 
generates control flow graphs (CFG) to facilitate further analyses.
Note that at this stage, the call graph is incomplete
due to the missing calling relationships of indirect calls.
\vspace{-1em}

\subsection{SSE-based Demand-driven Alias Analysis}
\label{section:alias}
\vspace{-1mm}
In this section, we describe the proposed demand-driven alias analysis.
We start with the definition of structured symbolic expressions (SSE), which is 
the basis of our approach.  Then, we explain
how to use SSE to find aliases of a given variable/pointer within
a basic block. We also show some running examples.
Finally, we illustrate how to implement intra- and inter-procedural alias analyses.


\begin{table}[t] \footnotesize
    \centering
    \vspace{-2.0mm}
    \setlength{\abovecaptionskip}{1mm} 
    \caption{Recursive definition of SSE.}
    \begin{tabular}{|ccc|}
    \hline
    $expr$ & $::=$ & $expr \ \lozenge _{b} \ expr\mid \lozenge _{u}expr\mid var$ \\
    $\lozenge _{b}$ & $::=$ & $+,-,*,/, \ll, \gg, ...$ \\
    $\lozenge _{u}$ & $::=$ & $\sim\ , !, ...$ \\
    $var$ & $::=$ & $\tau _{reg}\mid \tau _{mem} \mid \tau _{val}$ \\
    $\tau _{reg}$ & $::=$ & $r_{i}$ \\
    $\tau _{val}$ & $::=$ & $\left \{ Integer \right \}$ \\
    $\tau _{mem}$ & $::=$ & $load(expr) \mid store(expr)$ \\    
    \hline
    \end{tabular}
    \label{expr}
    \vspace{-7.0mm}
\end{table}

\vspace{-1em}
\subsubsection{Definition of SSE}
\label{sec:sse}
\vspace{-2mm}

SSE is a new notation that uses abstract memory model to represent 
aliases of a variable by encoding the nested provenance information at relevant program points. In
Table~\ref{expr}, we recursively defines an SSE expression. Note that since
our implementation is based VEX IR, our SSE definition is deeply influenced
by its design, in particular the basic statements and memory model. An SSE
could be any basic variable or the results of a binary/unary arithmetic
operation over basic variables. The basic variable can be either a
register (denoted as $\tau _{reg}$), a primitive immediate value (denoted as
$\tau _{val}$), or a memory access (denoted as $\tau _{mem}$). The
memory access can be a value loaded from memory (denoted as $load
(expr)$) or a value stored to memory (denoted as $store(expr)$). Here, the
$expr$ is a pointer.

We have implemented SSE based on \textsl{Claripy}, a popular
SMT solver engine used in angr.
\textsl{Claripy} provides a unified way to represent concrete and symbolic expressions.
It supports all common arithmetic operations (+, -, *, /, etc.)
with an extensible interface~\cite{ClaripyAPI}.
We leveraged this interface to
add the support of the memory \textit{load} and \textit{store} operations.

\begin{figure}[h]
\flushleft
\vspace{-3.0mm}
\includegraphics[width=0.70
\columnwidth]{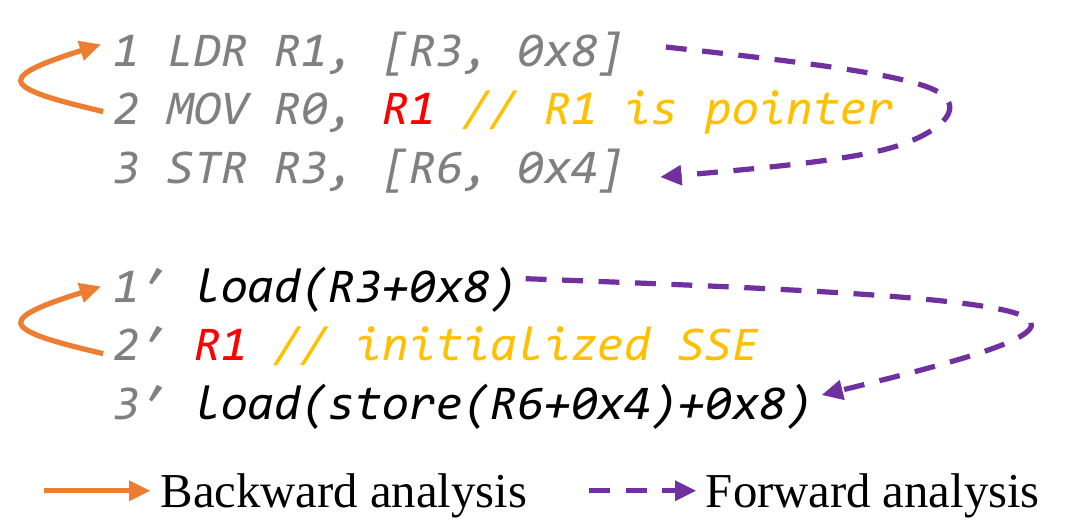}
\setlength{\abovecaptionskip}{1.0mm}
\caption{An intuitive simple example to illustrate SSE-based alias analysis}
\label{fig:sse}
\vspace{-5.0mm}
\end{figure}

\paragraph{An intuitive example.}
Now, we use an intuitive example to explain how we leverage SSE to find
aliases to a given pointer. In the code snippet in Figure~\ref{fig:sse}, there are three instructions. Our goal is to find all aliases
to \texttt{R1} in line 2, which is known to be a pointer. First, we
initialize the alias set with \texttt{R1} in line $2^\prime$. By looking backward, we
find a definition of \texttt{R1} in line 1, which gets its value by loading
from memory \texttt{R3+0x8}. Therefore, by replacing \texttt{R1} with \texttt
{load(R3+0x8)}, we add \texttt{load(R3+0x8)} in line $1^\prime$ to the alias set. Now,
if we look forward from line 1, we find a usage of \texttt{R3} in line 3.
That is, the value of \texttt{R3} is stored in the memory \texttt
{R6+0x4}. Therefore, by replacing \texttt{R3} with \texttt{store(R6+0x4)}
in \texttt{load(R3+0x8)}, we add \texttt{load(store(R6+0x4)+0x8)} in line
$3^\prime$ to the alias set. By doing so back and forth
iteratively along the define-use and use-define chain, we can eventually
reach a fixpoint of the alias set for the given pointer. The full SSE update
rules within basic blocks and
the algorithm for intra- and inter-procedural analyses are explained in the
following sections.

By using an SSE to encode pointer provenance information,
complex data structures (\eg~x.y.z) can be naturally traced.
Due to the on-demand feature,
it does not need to analyze the program as a whole.
Instead, we can focus on a particular variable and only compute over
relevant instructions. 
Therefore, our approach scales well for complex programs.







\begin{table*}[htb]
  \centering
  \setlength{\abovecaptionskip}{1.0mm}
  \caption{Update rules for structured symbolic expressions}
  \begin{adjustbox}{max width=0.80\textwidth}
  \begin{tabular}{ll}
    \toprule
    \multicolumn{2}{c}{\textbf{SSE update with define-use chain}} \\ \midrule
    
    $(1)ri = rj\xrightarrow{rj}expr.replace\left ( rj,ri \right ) $ 
    & $(2) ri = Binop\left ( rn,rm \right )\xrightarrow{rn \lozenge _{b} rm}expr.replace\left ( rn \lozenge _{b} rm,ri \right ) $ \\ 
    
    $(3) ri = ITE\left ( rj,rn,rm \right )\xrightarrow[rj]{rn}expr.replace\left ( rn,ri \right ) ^{\S} $ 
    & $(4) ri = ITE\left ( rj,rn,rm \right )\xrightarrow[!rj]{rm}expr.replace\left ( rm,ri \right ) ^{\S} $  \\ 
    
    $(5) ri = Load\left ( rj \right )\xrightarrow{load\left ( rj \right )}expr.replace\left ( load\left ( rj \right ),ri \right ) $ &
    $(6) Store\left ( ri \right ) = rj\overset{rj}{\rightarrow}expr.replace\left ( rj,store\left ( ri \right ) \right ) $ \\ \addlinespace%
    
    $(7) ri = Load\left ( rj \right )\xrightarrow{store\left ( rj \right )}expr.replace\left ( store\left ( rj \right ),ri \right ) $ & \\ \addlinespace%
    
    \multicolumn{2}{c}{\textbf{SSE update following use-define chain}} \\ \midrule
    
    $(8) ri = rj\xrightarrow{ri}expr.replace\left ( ri,rj \right ) $
    & $(9) ri = Binop\left ( rn,rm \right )\xrightarrow{ri}expr.replace\left ( ri,rn \lozenge _{b} tm \right ) $ \\ 
    
    $(10) ri = ITE\left ( rj,rn,rm \right )\xrightarrow[rj]{ri}expr.replace\left ( ri,rn \right ) ^{\S} $ 
    & $(11) ri = ITE\left ( rj,rn,rm \right )\xrightarrow[!rj]{ri}expr.replace\left ( ri,rm \right ) ^{\S} $  \\ 
    
    $(12) ri = Load\left ( rj \right )\xrightarrow{ri}expr.replace\left ( ri,load\left ( rj \right ) \right ) $
    & $(13) Store\left ( ri \right ) = rj\xrightarrow[*]{load\left ( ri \right )}expr.replace\left ( load\left ( ri \right ) ,rj  \right ) $ \\ \addlinespace%
    
    \multicolumn{2}{c}{\textbf{SSE kills}} \\ \midrule
    $(14) ri =rj\xrightarrow{ri}expr.kill\left ( \right ) $ 
    & $(15) Store\left ( ri \right )=rj\xrightarrow[+]{store\left ( ri \right ) \; or \; load\left ( ri \right )}expr.kill() $ \\ 
    
    \bottomrule
  \end{tabular}
  \end{adjustbox}
\vspace{-2mm}
\flushleft
\scriptsize{
\S: The statement $ri=ITE(rj,rn,rm)$ denotes that if $rj$ is true, $ri=rn$, otherwise, $ri=rm$. \quad
*: Existing $load(ri)$ in $expr$ occurs after the newly encountered $store(ri)$ statement. \\
+: Existing $store(ri)$/$load(ri)$ in $expr$ occurs before the newly encountered $store(ri)$ statement.
}

  \label{tab:updating}
  \vspace{-5.0mm}
\end{table*}

\begin{algorithm}[!t]
\small
\caption{Alias update within a basic block}
\label{alg:algorithm1}

\begin{algorithmic}[1]

\Procedure{TraceBlock}{$PRE_{f}$, $SUC_{b}$}
    \State $TARGET_{f}$, $TARGET_{b}$ $\leftarrow$
    \Call{GetCurrentTarget}{ }
    \State $IN_{f} \leftarrow PRE_{f}$ $\cup$ $TARGET_{f}$ 
    \State $IN_{b} \leftarrow SUC_{b}$ $\cup$ $TARGET_{b}$
    \Do
        \State $NEW_{f}$, $NEW_{b}$ $\leftarrow$ \Call{ForwardUpdate}{$IN_{f}$}
        \State $OUT_{f} \leftarrow OUT_{f} \cup NEW_{f}$        
        \State $IN_{b} \leftarrow IN_{b} \cup NEW_{b}$
        \State $NEW_{f}$, $NEW_{b}$ $\leftarrow$ \Call{BackwardUpdate}{$IN_{b}$}
        \State $OUT_{b} \leftarrow OUT_{b} \cup NEW_{b}$ 
        \State $IN_{f} \leftarrow NEW_{f}$
        \State $IN_{b} \leftarrow \emptyset $
    \doWhile{new aliases can be generated} 
    \State \Return $OUT_{f}$, $OUT_{b}$
\EndProcedure
\end{algorithmic}
\end{algorithm}

\vspace{-4mm}
\subsubsection{Alias Update Algorithm within a Basic Block}
\label{section:update}
\vspace{-2mm}

In this section, we describe how to find aliases for a given pointer within
a basic block.
Our algorithm, called \textsc{TraceBlock}, is shown in Algorithm~\ref{alg:algorithm1}.
As mentioned before, for a given pointer,
we search both forwards and backwards to enrich the alias set.
As such, for each basic block, we maintain two different sets of aliases
for forward and backward results respectively.
When running \textsc{TraceBlock} over a block,
it takes two parameters -- one forward alias set from all $current block$'s predecessors (denoted as $PRE_{f}$)
and one backward alias set from all $current block$'s successors (denoted as $SUC_{b}$).
At the initial basic block,
$PRE_{f}$ and $SUC_{b}$ are empty.
In line 2, \textsc{GetCurrentTarget} obtains
the manually specified target pointers of the current basic block (\eg~the taint source).
Then, in line 3, $IN_{f}$ is initialized to be the union of $PRE_{f}$ and $TARGET_{f}$.
In line 4, $IN_{b}$ is initialized to be the union of $SUC_{b}$ and $TARGET_{b}$.
Here, $IN_{f}$ contains all the SSEs needing forward tracking and
$IN_{b}$ contains all the SSEs needing backward tracking.
Forward and backward tracking are implemented via \textsc{ForwardUpdate} and \textsc{BackwardUpdate} respectively.
\textsc{ForwardUpdate} takes $IN_{f}$ as input and generates new forward
SSEs, which are merged into $OUT_{f}$.
\textsc{ForwardUpdate} may also return new backward SSEs, which are merged into $IN_{b}$
for further processing.
\textsc{BackwardUpdate} takes $IN_{b}$ as input and generates new backward
SSEs, which are merged into $OUT_{b}$.
\textsc{BackwardUpdate} may also return new forward SSEs, which are merged into $IN_{f}$
for further processing.
The algorithm runs iteratively and terminates when
no new alias SSE can be generated (line 5-12).
The output of the algorithm, $OUT_{f}$ and $OUT_{b}$,
are used as parameters for tracking succeeding basic blocks and preceding 
basic blocks respectively (see Section~\ref{sec:intra} for more details).
In the following, we detail the forward update and backward update processes respectively.




\paragraph{Forward update.}
During forward analysis, \sys traverses instructions
following the instruction flow.
In the simplest form,
for a live variable $v$ in the current SSE, 
if $v$ is on the RHS (right hand side) of an assignment statement $d=v$ (\ie~$v$ is used
in a define-use chain),
a new SSE is generated by replacing the variable $v$ in the original SSE
with the variable $d$. 
Therefore, the new SSE becomes an alias to the old one, 
and the new SSE needs to continue to be tracked forwards.
The full SSE update rules following the define-use chain are listed
in entries 1-7 in Table~\ref{tab:updating}.
In the table, each entry is represented as
$statement \xrightarrow[rm]{rn} expr.replace(rn,rj)$,
where $statement$ is the encountered instruction and $expr$ is the tracking SSE.
In the $statement$, 
if $rm$ is true and $rn$ exists in $expr$,
then $rn$ should be replaced with $rj$. 
Note that this process should be conducted over all the live variables in the
current SSE.

As mentioned before, a forward update operation following the define-use chain can also
generate SSEs needing backward tracking.
Specifically,
one type of assignment statement is in the form of $store(d)=v$ and $v$ is a pointer.
For such type, in addition to continuing forward tracking, 
the new SSE needs to be tracked backwards also to find the definitions
of the store address, \ie~$d$, which is because
once such a definition is found, a new alias can be generated.

\paragraph{Backward update.}
During backward analysis, \sys traverses instructions
following the instruction flow reversely.
In the simplest form, for a live variable $v$ in the current SSE,
if $v$ is on the LHS (left hand side) of an assignment statement $v=u$ (\ie~$v$ is defined in a use-define chain),
a new SSE is generated by replacing the variable $v$ in the original SSE with the variable $u$.
Therefore, the new SSE becomes an alias to the old one.
Now that $u$ is live in both directions, 
the new SSE needs to be tracked both backwards and forwards
to find the definitions and uses of the variable $u$.
The full SSE update rules following the use-define chain are listed
in entries 8-13 in Table~\ref{tab:updating}.


Different from forward tracking, in backward tracking, we should not only find
definitions of a variable $v$ (\ie~$v=u$ as explained before),
but also find its uses (\ie~$d=v$).
If a use is found, we follow entries 1-6 in Table~\ref{tab:updating} except for entry 7 to
update new SSEs.
However, 
the new SSE can only be tracked forward, not backward,
because the variable $d$ used to update the original SSE is live only in forward direction.
One type of assignment statement is in the form of $store(d)=v$ and $v$ is a pointer.
Note that for such type, in addition to forward tracking, the new SSE needs to be tracked backwards also, to find the definition of store address (\ie~$d$).

\paragraph{Conditions to kill an SSE.}
Each variable has its own live scope.
The liveness of the corresponding register (\ie~$\tau _{reg}$) is killed 
if we encounter a register assignment statement.
If this happens in forward/backward analysis, the whole SSE with this register is killed.
This corresponds to entry 14 in Table~\ref{tab:updating}.
Similarly, the liveness of the corresponding memory access (\ie~$\tau _{mem}$) is killed
if we encounter a $store$ statement.
If this happens in forward analysis, the whole SSE with this variable is killed.
This corresponds to entry 15 in Table~\ref{tab:updating}.

\begin{figure}[h]
\flushleft
\vspace{-3.0mm}
\includegraphics[width=0.7
\columnwidth]{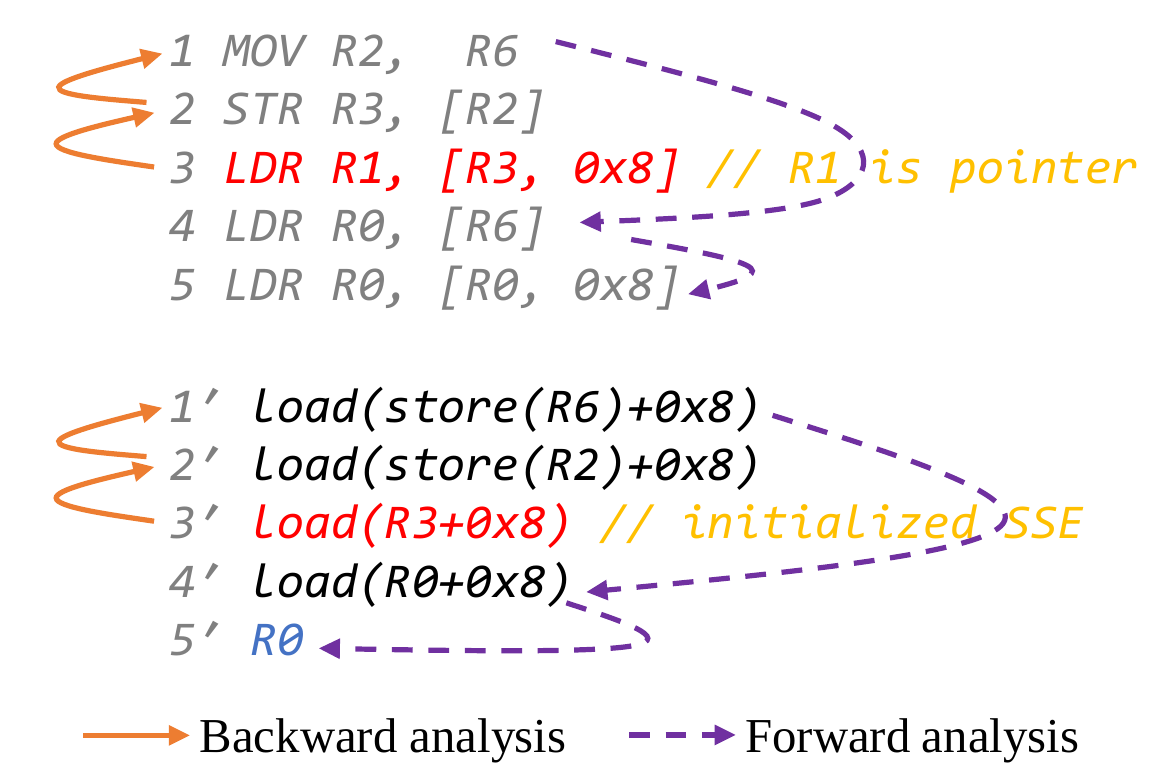}
\setlength{\abovecaptionskip}{1.0mm}
\caption{A complex example to illustrate SSE-based alias analysis.} 
\label{fig:alias}
\vspace{-5.0mm}
\end{figure}

\paragraph{A non-trivial running example.} Following 
Algorithm~\ref{alg:algorithm1}, we are able to find alias information 
for more complex programs. We show such an example in Figure~\ref{fig:alias},
in which \texttt{R1} in line 3 and \texttt{R0} in line 5 are actually aliases
to each other.

Assuming that the analysis begins at line 3, in which \texttt{R1}
is load from the address \texttt{R3+0x8}.
To find aliases of \texttt{R1},
\sys first initializes an SSE as \texttt{load(R3+0x8)} in line $3^\prime$. In the first
 round of forward analysis, it does not find any use of \texttt{R3}, so it
 begins backward analysis. Searching backwards, \sys finds a use of \texttt{R3} in line 2. So we 
 replace \texttt{R3} in \texttt{load(R3+0x8)} with \texttt{store(R2)} following rule 6 in Table~\ref{tab:updating} and then we get a new alias SSE \texttt{load(store(R2)+0x8)}. Continuing backward tracking, \sys finds a definition
 of \texttt{R2} in line 1. Then it follows rule 8 to replace \texttt{R2} with \texttt{R6}, yielding a new
 alias SSE \texttt{load(store(R6)+0x8)}.
 This ends the first iteration and
 both new alias
 SSEs need to be tracked backwards and forwards for a second round. We
 start with forward analysis for \texttt{load(store(R6)+0x8)} in line 1. \sys
 quickly finds a use of \texttt{load(R6)} in line 4. Following rule 7, \sys
 replaces \texttt{store(R6)} with \texttt{R0} and gets a new alias \texttt{load(R0+0x8)} in line $4^\prime$. Finally, following rule 5 in line $5^\prime$, \sys
 gets \texttt{R0} as a new alias. This concludes that \texttt{R1} in line 3
 is an alias to \texttt{R0} in line 5.





\subsubsection{Intra- and Inter-procedural Alias Analyses}
\label{sec:intra}
\vspace{-2mm}

We have described how to find aliases within a basic block.
In this section, we introduce intra- and inter-procedural alias analysis,
which we summarize in Algorithm~\ref{alg:algorithm3}.
The procedure for finding aliases for a function is called \textsc{AnalyzeFunction},
which takes the CFG of the function as input.
First, a $WorkList$ of basic blocks is obtained via postorder traversal of the CFG (line 2),
in which a block is visited after all its successor blocks have been visited. 
Then, the procedure \textsc{FindAlias} is used
to actually analyze each block.
We use the same procedure over the $WorkList$ forwards and backwards,
until no new alias can be found.

Inside \textsc{FindAlias},
\texttt{TraceBlock} is applied to analyze aliases within each basic block $BB$ (line 16).
Then the results are merged to the successors and predecessors of the current $BB$ (line 17-18).
However, if the current basic block is a callsite,
we need to incorporate inter-procedural analysis.
Note that in our design, a call instruction is treated as a separated basic block,
which may differ from the definition of basic blocks in other tools.
Specifically, \textsc{FindAlias}
collects modifications (MOD) and references (REF) to memory locations 
that are accessed directly or indirectly through parameters, return values, or global pointers in the callee (line 11-14).
In other words,
we summarize the modifications and references to memory locations by callee.
To do so, at first, 
the algorithm checks the SSEs in the forward alias set $PRE_{f}$
and backward alias set $SUC_{b}$ of the callsite $BB$, 
extracts the root pointer of each SSE 
(\eg~the root pointer of \texttt{load(R0+0x8)} is \texttt{R0}),
and stores them in a set called $PTRs$ (line 12).
Then, we use a transfer function (\textsc{TransferFun})~\cite{cheng2000modular} to actually get $MOD$ and $REF$.
Our implementation of transfer function is quite standard so we omit the details.
Finally, with $MOD$ and $REF$,
\textsc{FindAlias} updates SSEs in $PRE_{f}$ and $SUC_{b}$ of 
the callsite $BB$ with the definitions in $MOD$ and $REF$ (line 14).

Back to \textsc{AnalyzeFunction},
after using \textsc{FindAlias} iteratively to reach a fixpoint,
the results of the current function are
merged to the caller.
Specifically, SSEs associated with arguments or global pointers 
from the entry block's $OUT_{b}$ are merged 
into $SUC_{b}$ of the caller's call site (line 7),
and SSEs associated with return value from the exit block's $OUT_{f}$
are merged into $PRE_{f}$ of caller's return site (line 8).

\begin{algorithm}[!t]
 \small
\caption{Intra- and inter-procedural alias analysis}
\label{alg:algorithm3}
\begin{algorithmic}[1]

\Procedure{AnalyzeFunction}{$CFG$} 
    \State $WorkList \leftarrow Postorder(CFG)$
    \Do
        \State \Call{FindAlias}{$WorkList.reverse()$} \Comment{forward} 
        \State \Call{FindAlias}{$WorkList$}           \Comment{backward}
    \doWhile{new aliases can be generated}
    \State \Call{Merge}{$Caller.CallSite$, $EntryBB.OUT_{b}$}
    \State \Call{Merge}{$Caller.ReturnSite$, $ExitBB.OUT_{f}$}


\EndProcedure

\Procedure{FindAlias}{$WorkList$} 
    \For{$BB \in WorkList$}
        \If{$BB$ is a callsite}
            \State $PTRs \leftarrow $ \Call{GetRootPtr}{$PRE_{f}$, $SUC_{b}$}
            \State $MOD, REF \leftarrow$ \Call{TransferFunc}{$PTRs$, $callee$}
            \State \Call{UpdateContext}{$PRE_{f}$, $SUC_{b}$, $MOD$, $REF$}
        \Else
            \State $OUT_{f}, OUT_{b} \leftarrow $\Call{TraceBlock}{$PRE_{f}$, $SUC_{b}$}
            \State \Call{Merge}{$BB.successors, BB.OUT_{f}$}
            \State \Call{Merge}{$BB.predecessors, BB.OUT_{b}$}
        \EndIf
    \EndFor
\EndProcedure


\end{algorithmic}
\end{algorithm}

\subsection{Indirect Call Resolution}
\label{section:recovery}
\vspace{-2mm}

\sys resolves indirect calls based on dependencies between the aliases of indirect call targets
and the aliases of function pointer references (or function table pointers).
First, \sys identifies all the indirect call instructions
by traversing through the disassembled code.
\sys treats all branch instructions
with a register as operand to be indirect calls (e.g., \texttt{blx r0} in ARM
and \texttt{jalr \$t9}) in MIPS).
\sys then uses IDA Pro to filter call instructions whose targets can be recognized.
The remaining ones were treated as unresolved indirect calls.
Since almost all the call instructions in MIPS use a register as operand,
MIPS programs have a considerably larger number of indirect calls 
than ARM programs (see Table~\ref{tab:icall}).


Second, \sys finds all address-taken functions by scanning both the code segment and data segment, and locates all the references to the address-taken functions.
There are mainly two ways to reference a function.
A \emph{function pointer} which contains the address of the function,
can be used directly as an operand in an indirect call.
We denote this kind of references as $fptr$ and show an example in line 2
of Listing~\ref{indirectcall}.
A \emph{function table pointer} which contains the address to
a function table, can be used indirectly (\ie~by deferencing the function table first) to get the real function address.
We denote this kind of references as $dptr$ and show an example in line 7
of Listing~\ref{indirectcall}.
In this example, an entry in the table has two fields -- a string pointer
to the function name and the actual function pointer.



\begin{lstlisting}[language=C, label=indirectcall, basicstyle=\scriptsize\ttfamily, xleftmargin=1.5em,framexleftmargin=2em, caption={Indirect call resolution example.}, escapeinside=``]
void (*fun_ptr)() = &fun; //address-taken function
`\textbf{fptr = fun\_ptr;}`
x.y.z = fptr;
fp1 = x.y.z;
`\textbf{call fp1;}`  //indirect call

`\textbf{dptr = 0x92C44;}`  //function table address
while ( strcmp(*dptr, name) ){ 
  dptr += 0x8;
  ...
}
fp2 = *(dptr+0x4); 
`\textbf{call fp2;}`  //indirect call
\end{lstlisting}
\vspace{-2mm}


Third, \sys performs alias identification. Our goal
is to find dependence between the indirect call targets (\eg~\texttt{fp1} and \texttt{fp2} in line 5 and 13)
and the original references (\eg~\texttt{fptr} and \texttt{dptr} in line 2 and 7).
\sys back tracks the indirect call targets
to find its all aliases.
At the same time, \sys finds the aliases of
$fptr$ and $dptr$ through both forward and backward analyses.
Then, \sys collects all the alias SSEs in the entry block and exit block of every function,
checks the data dependency between aliases of $fptr/dptr$ and aliases of indirect call target,
and updates them.
We denote the alias SSEs of the indirect call targets (\eg~\texttt{fp1} and
\texttt{fp2} in the listing) as \cexpr (call target expression), 
and the alias SSEs of the $fptr$ or $dptr$ as \pexpr (pointer expression).




Finally, \sys matches the aliases to eventually resolve indirect calls.
In the simplest form, a \cexpr is an alias to \pexpr.
\sys can calculate the call target directly from \cexpr.
This often indicates an indirect call by $fptr$.
If the \cexpr can be represented by $load(dptr+i*stride)$, where the $dptr$ is the
address of a function table,
the $stride$ is a constant, and $i$ is an index,
it strongly indicates a function table deference.
\sys reads values from addresses $dptr+i*stride$, 
where the index $i$ starts at 0 and increases by 1 at a time.
It terminates when the retrieved value is greater than the maximum address of the function table.
If the returned value is a legitimate function address, \sys adds it to the set of indirect call targets. 


For the indirect call targets that cannot be accurately 
traced to a specific $fptr$ or $dptr$,
we indirectly find their dependencies.
Our observation is that
$fptr$ and $dptr$ are often stored in a multi-level data structure
whose root pointer is a global pointer (denoted as $gptr$).
Therefore, we often find $store(gptr)$
or $store(gptr)$ in \pexpr.
To indirectly call the target, the program also needs to refer to 
the same global pointer before the callsites.
If the \cexpr can be simplified to $load(gptr)$ where the same $gptr$ is used in the \pexpr,
the indirect call target is immediately recovered by using $fptr$.
If the \cexpr can be simplified to $load(load(gptr)+i*stride)$
where the same $gptr$ is used in \pexpr,
by replacing $load(gptr)$ in \cexpr with $dptr$ (since $dptr$ has an alias SSE $load(gptr)$),
a new SSE $load(dptr+i*stride)$ can be generated.
Then, \sys resolves the indirect call targets using $load(dptr+i*stride)$.

\vspace{-5mm}
\subsection{Taint-style Vulnerability Check}
\label{section:taint}


The taint-style vulnerability is caused by the lack of security sanitization on 
security-sensitive sinks whose data can be propagated from attacker-controlled sources.
Our goal is to track the tainted data from sources through the proposed alias analysis,
examine the constraints of the tainted data, 
and determine whether the tainted data can be propagated to sinks.
Note that our analysis is performed on the improved CFG where
indirect calls have been recovered.

\paragraph{Taint Source and Sink.}
In embedded firmware analysis, \sys treats the data received from network as taint sources.
In C code, sources include \textit{recv/recvfrom} and other library functions,
such as \textit{read/fgets}, which may read data from a local file or network.
For these functions, \sys performs a backward data flow analysis on its file descriptor parameter.
If the parameter data depends on a specific file, we remove it from the source list.
Particularly, \sys identifies function \textit{getenv} as a source.
The sinks include string copy library functions (e.g. \textit{strcpy, memcpy}) and
command execution library functions (e.g. \textit{system, popen}).
Table~\ref{tab:source} in Appendix~\ref{source-sink} shows all sources and sinks used in \sys.

\paragraph{Library Function Summaries.}
In taint analysis, the tainted data may be propagated to library functions.
\sys adopts ``function summaries”, which describes what variables are tainted 
and how these tainted variables are propagated within a function, to handle common string functions.
We have implemented summaries for 29 common functions from the Standard C Library (Table~\ref{tab:lib-tran} in Appendix), such as the function \texttt{strcpy(*dest,*src)}, in which the taint is propagated from argument \texttt{src} to argument \texttt{dest}.



\paragraph{Taint propagation.}
The taint analysis is implemented based on the SSE-based demand-driven alias analysis.
Therefore, taints are propagated along with the alias analysis.
For a given source, 
\sys taints relevant pointers depending on the semantic of the source function,
which can either be function parameters or a return value.
Then, \sys tracks the tainted SSE to find its aliases and
propagates the taint to another following arithmetic operations (\ie \texttt{a=b+1}), 
library functions listed in Table~\ref{tab:lib-tran} in Appendix~\ref{lib-summar}, and loop copy functions.
This is quite standard in all related works.
During taint analysis, \sys also collects constraints for the tainted SSE (\eg~\texttt{x<4} where \texttt{x} is tainted).
Whenever the tainted SSE reaches a sink, 
both the sink's information (e.g., sink's function name, buffer size)
and the corresponding tainted SSE with its constraints are collected and passed to the taint-style vulnerability check module for further analyses.

In forward analysis, when the tracked SSE contains the parameters of the callee, 
or when the root pointer of the SSE is a global pointer,
\sys goes into the callee to continue the taint analysis. 
As an optimization, for each analyzed function, \sys creates a function summary that describes tainted input data sets and the corresponding tainted output data sets.
When \sys encounters the same function, if the tracked SSE is in the tainted input data sets, the taint is propagated directly based on the summary.
Otherwise, \sys analyzes the function again and complements the existing summary.

However, not all aliases should be tainted following the alias analysis.
For example, the buffer pointer tainted in \textit{recv} should not
propagate backwards when \textit{recv} has not invoked yet.
\sys determines whether a new alias SSE should to be
tainted based on what we call taint-trigger points.
The taint-trigger points are a class of callsites,
which satisfies the property that the callee's argument pointer is 
tainted after the callee is executed.
\sys tracks the taint-trigger point for each alias.
Whenever an alias is propagated over its taint-trigger point in forward analysis,
the alias is tainted.
In contrast, whenever an alias is propagated across its taint-trigger point in backward analysis,
the alias is untainted.
At sinks, we only consider tainted SSEs to detect taint-style vulnerabilities.


\paragraph{Constraint checking.}
Constraint checking is the last step to find potential bugs.
We follow a quite standard approach.
Specifically, at sinks,
\sys checks the associated tainted SSE parameters.
If it contains a symbolic constraint (\ie~\texttt{str\_len < s}),
\sys does not report an alert.
Here, the \texttt{str\_len} is the length of tainted string.
Otherwise, \sys solves the constraints to get its minimum value \texttt{min\_len}.
For memcpy-like functions,
if they copy data to stack buffers, \sys retrieves the address of the destination buffer.
If \texttt{min\_len} is larger than the distance from the buffer address to
the top of the stack or there is no constraint on the tainted SSE,
an alert is raised.
Likewise, for the command execution library functions (e.g. \textit{system}),
\sys reports an alert if there is no constraint on the tainted SSE.
\vspace{-1em}




\begin{table*}[t]
    \centering
    \setlength{\abovecaptionskip}{1.0mm}
    \caption{\label{tab:vul} Alerts produced by \sys for 35 samples} 
    \begin{adjustbox}{max width=0.98\textwidth}
    \begin{tabular}{lccccrrrrr} 
        \toprule    
        \textbf{Vendor} & \textbf{ID} & \textbf{Firmware Version} & \textbf{Arch} & 
        \textbf{Binary} & \textbf{Size (KB)} & \textbf{Ana. Func} & \textbf{Tainted Sinks} & \textbf{Alerts} & \textbf{Time (s)} \\ 
        \midrule 

        \multirow{2}{*}{\textbf{Cisco (2)}} 
        & 1 & RV320\_v1.4.2.20 & MIPS64 & ssi.cgi & 1,820 & 1,567 & 1450 & 335 & 634.64 \\
        & 2 & RV130\_v1.0.3.44 & ARM32  & httpd   & 612   &  796  &  402 & 150 & 60.96  \\ \addlinespace%

        \multirow{2}{*}{\textbf{D-Link (7)}}
        & 3 & DIR-825\_B\_2.10 & MIPS32 & httpd & 531 & 447 & 198 & 36 & 95.02 \\ 
        & 4 & DAP-1860\_A1\_B03 & MIPS32 & uhttpd & 1,129 & 1,030  & 106 & 12 & 97.02 \\ \addlinespace%
        
        \multirow{2}{*}{\textbf{TRENDnet (2)}}
        & 10 & TEW632BRP\_1.010B32 & MIPS32 & httpd & 314 & 315  & 149 & 26   & 40.08  \\
        & 11 & TEW827DRU\_2.04B03 & MIPS32 & ssi & 998 & 622  & 289 & 142  & 58.20 \\ \addlinespace%
        
        \multirow{2}{*}{\textbf{NETGEAR (17)}}
        & 12 & R7800\_v1.0.2.32 & ARM32 & net-cgi & 542 & 1,286  & 291 & 119 & 88.91 \\
        & 13 & R8000\_v1.0.4.4 & ARM32 & httpd & 1,508 & 1,088  & 428 & 36  & 167.48  \\ \addlinespace
        
        \multirow{1}{*}{\textbf{TP-Link (3)}}
        & 29 & WR940NV4\_us.3.16.9 & MIPS32 & httpd & 1,691 & 3,481  & 225 & 31 & 343.16 \\ \addlinespace
        
        \multirow{1}{*}{\textbf{Tenda (4)}}
        & 32 & AC9V3.0\_v15.03.06.42 & MIPS32 & httpd & 2,039 & 1,201  & 172 & 84 & 433.62 \\
        
        \hline
        \textbf{Total (35)$^{\S}$} & - & - & - & - & - & 38,983 & 9,346 & 1,887 & 179.81$^{\dag}$ \\

        \bottomrule 
    \end{tabular} 
    \end{adjustbox}
\vspace{-2mm}
\flushleft
\scriptsize{
\S: The row labelled ``Total'' shows aggregated results for 35 samples.
Due to page limit, we only list 10 representative samples in the table.
The full list can be found in Table~\ref{tab:vul2} in Appendix. 
\quad \quad
\dag: The average execution time per sample.
}
    \vspace{-5.0mm}
\end{table*}

\begin{table}[ht]
    \centering
    \setlength{\abovecaptionskip}{1.0mm}
    \caption{\label{tab:sample} True positive evaluation by random sampling} 
    \begin{adjustbox}{max width=0.85\columnwidth}
    \begin{tabular}{l l c c c  } 
        \toprule    
        \textbf{ID} & \textbf{Model} & \textbf{Alerts} & \textbf{\# of Samples} & \textbf{\# of TP} \\
        \midrule 
        1 & Cisco RV320 & 335 & 34 &  32 \\

        2 & Cisco RV130 & 150 & 15 &  15 \\

        3 & D-Link DIR-825 & 36 & 4 &  3 \\

        4 & D-Link DAP-1860 & 12 & 2 &  2 \\

        10 & TRENDnet TEW632BRP & 26 & 3 &  2 \\

        11 & TRENDnet TEW827DRU & 142 & 14  & 7 \\

        12 & NETGEAR R7800 & 119 & 12 &  10 \\

        13 & NETGEAR R8000 & 36 & 4 &  4 \\

        29 & TP-Link WR940NV4 & 31 & 4 &  4 \\

        32 & Tenda AC9 & 84 & 8 &  7 \\

        \hline

        Total & 10 & 971 & 100 &  86 \\

        \bottomrule 
    \end{tabular} 
    \end{adjustbox}
\end{table}

\vspace{-2mm}
\section{Evaluation}
\vspace{-2mm}
We have implemented a prototype of \sys, which consists of about 24,000
LoC in Python.
We evaluated \sys from three aspects.
(1) How effective is it in uncovering real-world taint-style vulnerabilities in embedded firmware (Section~\ref{sec:eval:vul})?
(2) How effective is it in indirect call resolution? 
To what extent does the indirect call resolution play a role in improving vulnerability discovery (Section~\ref{section:icall})?
(3) How effective is it compared with the state-of-the-art tools (Section~\ref{section:comp2})?


\paragraph{Experiment setup.}
Our evaluation was conducted against 35 different firmware samples 
from six major vendors of network embedded systems:
Cisco, D-Link, NETGEAR, TRENDnet, TP-Link and Tenda.
These samples were manually downloaded from the vendor's official websites.
In order to compare with the existing tools head-to-head,
some samples are outdated. However,
as we will explain later, all the identified bugs were tested against
the latest firmware versions before reporting them to the vendors.
Table~\ref{tab:vul} summarizes the information about each sample,
including its vendor, firmware version, architecture,
the analyzed Internet-facing binary, its size, etc.
This sample set covers three architectures, ARM32, MIPS32 and MIPS64, which are
the mainstream architectures used in embedded devices.
Note that since our approach works based on VEX IR,
it can be easily extended to support other architectures.
All the experiments were conducted on a Ubuntu 18.04.4 LTS OS
running on PC with a 64-bit 8-core Intel(R) Core(TM) i7-8550U CPU and 24 GB RAM.

\vspace{-4mm}
\subsection{Vulnerability Discovery}
\label{sec:eval:vul}
\vspace{-2mm}

Our current prototype, \sys, checks unsafe data copy functions (\eg~\texttt{strcpy})
and command execution functions (\eg~\texttt{system}).
Therefore, \sys reports vulnerabilities related to buffer overflow or command injection.
We summarize the results in Table~\ref{tab:vul}.
In total, \sys reached 9,346 different sinks where there are tainted parameters
to the sink functions.
1,887 of them were identified as alerts
because no security sanitization was detected.
The average time to analyze each sample is about 180 seconds.


\paragraph{True-positive evaluation.} Given the large amount of raised alerts,
 it is important to understand how many of them are true positives. However,
 this would require substantial human efforts to manually craft a
 proof-of-concept (PoC) for each of them on real devices. Reverse-engineering
 each sample is also impractical because we have no ground truth of the
 indirect calls. Therefore, we adopted random sampling in an attempt to obtain an
 estimated true positive rate. Specifically, we acquired 10 physical devices
 and randomly sampled 100 alerts out of 971 that correspond to these 10
 devices. Then, we manually analyzed these alerts. An alert is confirmed as a
 true positive if (1) it matches a known vulnerability or (2) can be verified
 by successfully constructing a PoC on the physical device. Table~\ref{tab:sample} shows the results. Out of 100 alerts, we confirmed 86 true
 positives, including 13 known vulnerabilities and 73 successfully
 constructed PoCs. This indicates the high true positive rate of our
 approach. For the rest, which are false positives, we attribute them to
 either inaccurate indirect call resolution or failing to find security checks
 (see Section~\ref{sec:limitaiton} for details).

\paragraph{False negative evaluation.} To evaluate the false negative rate
 (\ie~the rate of failing to find a bug) of our prototype, we collected 42
 known vulnerabilities with exposed details from exploit-db and MITRE CVE for
 the 35 firmware samples in our dataset. \sys discovered 41 of them
 (see Table~\ref{tab:OneDay} in Appendix). The missing one (CVE-2019-6989) is a buffer
 overflow bug caused by unsafe copy using loop, which our tool does not check
 currently. We will
 extend our tool in the future to support more kinds of security
 checks at sink.

\paragraph{N-day and 0-day vulnerabilities.}
In addition to verifying alerts through sampling as mentioned before,
we have been working hard to verify more alerts.
We prioritize alerts related to physical devices that we have access to.
At the time of writing,
we have confirmed a total of 41 n-day vulnerabilities and 151 0-day vulnerabilities,
115 of which have been assigned with CVE/PSV numbers.
As mentioned before,
the 41 n-day vulnerabilities were matched
from exploit-db~\cite{exploitDB} or MITRE CVE~\cite{cve},
which correspond to 120 alerts.
In Table~\ref{tab:OneDay} of Appendix, we list how the identified 41 n-day vulnerabilities
are distributed among the public exposure IDs.
Note that one CVE ID may correspond to multiple alerts.
For example, CVE-2019-13278 describes a command injection vulnerability, which can be triggered at 33 different sinks.

To confirm 0-day vulnerabilities, we tried to craft PoCs against the latest
firmware versions of physical devices.
In total, we confirmed 151 0-day vulnerabilities,
including 38 command injection bugs and 113 buffer overflow bugs.
Table~\ref{tab:0day} in Appendix summaries these bugs and lists relevant CVEs/PSVs.
As mentioned before, some of the 35 samples are old-versioned for comparative evaluation,
therefore, we found that lots of alerts generated by \sys
have been fixed in the latest version.
These bugs were probably found by the vendor themselves,
therefore no public detail is available.
For example,
\sys produced 119 alerts in NETGEAR R7800 v1.0.2.32, but we only verified one
 0-day vulnerability in its latest version v1.0.2.68. By reverse-engineering
 both samples, we found that many bugs were fixed by replacing \texttt
 {strcpy} with \texttt{strncpy} and setting the string length parameter to a
 constant.
 

\begin{table}[htb] 
    \centering
    \setlength{\abovecaptionskip}{1.0mm}
    \caption{\label{tab:icall}Results of indirect call resolution} 
    \begin{adjustbox}{max width=\columnwidth}
    \begin{tabular}{llrrrrr} 
        \toprule   

        \multirow{2}{*}{\textbf{ID}} & \multirow{2}{*}{\textbf{Model}} & \textbf{All} & 
        \textbf{Resolved} & \textbf{I-Call} &  \textbf{\% of resolved } & \multirow{2}{*}{\textbf{Time (s)}} \\

         & & \textbf{I-Calls} & \textbf{I-Calls} & \textbf{targets} & \textbf{I-Calls}  \\

        \midrule 

        1 & Cisco RV320 & 638 & 620  & 794 & 97.2\% & 288.04  \\
        2 &   Cisco RV130 & 17 & 14    & 475  & 82.4\% & 27.83  \\ 

        3 & D-Link DIR-825 & 82 & 80  & 239  &  97.5\% & 33.44  \\ 
        4 & D-Link DAP-1860 & 27 & 21 & 327 &  77.8\% & 37.15 \\ 
        
        10 & TRENDnet TEW632BRP & 43 & 41 & 182 &  95.3\% & 20.40  \\
        11 & TRENDnet TEW827RU & 51 & 48  & 381 &  94.1\% & 25.56  \\ 
        
        12 & NETGEAR R7800 & 17 & 14  & 692 &  82.3\% & 40.27  \\
        13 & NETGEAR R8000 & 3 & 2 & 491 &  66.6\% & 40.35  \\
        
        29 & TP-Link WR940NV4 & 389 & 309 & 653 &  79.4\% & 409.10  \\ 
        
        32 & Tenda AC9V3.0 & 88 & 65 & 286 & 73.9\% & 204.55  \\
        \hline
        \textbf{Total (35)$^{\S}$} & - & 12,446 & 12,022 & 14,920 & 96.6\% & 95.55$^{\dag}$ \\
        \bottomrule 
    \end{tabular} 
    \end{adjustbox}
\vspace{-3mm}
\flushleft
\scriptsize{
\S: The row labelled ``Total'' shows aggregated results for 35 samples.
Due to page limit, we only list 10 representative samples in the table.
The full list can be found in Table~\ref{tab:icall2} in Appendix. 
\quad
\dag: The average execution time per sample.
}
    \vspace{-5.0mm}
\end{table}

\begin{table*}[t!]
    \centering
    \setlength{\abovecaptionskip}{1.0mm}
    \caption{\label{tab:compOne} The impact of indirect call resolution on vulnerability discovery.}
    \begin{adjustbox}{max width=0.85\textwidth}
    \begin{tabular}{lrrrr|rrrr}
        \toprule    
        \multirow{2}{*}{\textbf{Model}} & \multicolumn{4}{c}{\textbf{w/o I-Call Resolution}} & \multicolumn{4}{c}{\textbf{I-Call Resolution}} \\
        & \textbf{Ana. Block} & \textbf{Tainted Block} & \textbf{Tainted Sink} & \textbf{Alerts}
        & \textbf{Ana. Block} & \textbf{Tainted Block} & \textbf{Tainted Sink} & \textbf{Alerts} \\
        \midrule 
        \textbf{Cisco RV320} & 74,778 & 976 & 98 & 16 & 89,915 & 16,307 & 1,450 & 335 \\
        \textbf{Cisco RV130} & 26,210 & 1,234 & 87 & 2 & 30,919 & 4,585 & 402 & 150 \\
        \textbf{D-Link DIR-825} & 19,661 & 1,994 & 161 & 18 & 21,169 & 2,553 & 198 & 36 \\
        \textbf{D-Link DAP-1860} & 34,741 & 2,303 & 100 & 8 & 34,881 & 2,326 & 106 & 12 \\

        \textbf{TRENDnet TEW632BRP} & 10,756 & 1,556 & 126 & 13 & 11,672 & 1,953 & 149 & 26 \\

        \textbf{TRENDnet TEW827RU} & 21,348 & 2,788 & 286 & 142 & 22,714 & 2,860 & 289 & 142 \\

        \textbf{NETGEAR R7800} & 29,640 & 2,382 & 118 & 8 & 32,262 & 5,217 & 291 & 119 \\

        \textbf{NETGEAR R8000} & 46,053 & 1,997 & 166 & 0 & 56,907 & 7,373 & 428 & 36 \\

        \textbf{TP-Link WR940NV4} & 59,786 & 1,987 & 50 & 1 & 69,759 & 7,258 & 225 & 31 \\

        \textbf{Tenda AC9V3.0} & 33,612 & 2,001 & 42 & 0 & 42,490 & 6,056 & 172 & 84 \\

        \hline
        \textbf{Total (10)} & \textbf{356,585} & \textbf{19,218} & \textbf{1,234} & \textbf{208} 
        & \textbf{412,688} & \textbf{56,488} & \textbf{3,710} & \textbf{971} \\
        \bottomrule 
    \end{tabular} 
    \end{adjustbox}
\end{table*}

\vspace{-2mm}
\subsection{Indirect Call Resolution and Its Importance}
\label{section:icall}

\vspace{-3mm}
\paragraph{Indirect call resolution.}
Table~\ref{tab:icall} shows the results of indirect call resolution over
the 35 samples.
The targets of our analysis are the binaries from Table~\ref{tab:vul}.
We first identified all the callsites of indirect calls in the binaries and then
attempted to find the called targets.
Note that IDA pro already has some limited indirect call resolution capability.
If IDA pro was able to resolve an indirect call,
the corresponding callsite was removed.
For each sample, the number of indirect callsites is denoted as \textbf{All I-Calls}.
If at least one target was found, we mark it as resolved.
We denote this number as \textbf{Resolved I-Calls}.
Our approach was able to resolve 12,022 indirect callsites (out of 12,446)
and the average execution time per sample is about 96 seconds.
We also calculated the total number of target functions called at indirect callsites during this process,
which is listed in the column \textbf{I-Call targets}.
In total, our tool added 14,920 target functions that are called indirectly,
which substantially improved taint-style vulnerability discovery.
Since we do not have the ground truth for indirect call targets,
we cannot accurately evaluate the failed resolutions.
However, by reverse-engineering some samples, we did
find some missed targets because some address-taken functions were not recognized.
Interestingly, our tool also resolved some indirect call targets to be NULL.
It turned out that the firmware later correctly performs checking
before dereferenceing them.
Therefore, our resolution was correct.


\paragraph{Importance of indirect call resolution.}
We analyzed the same 35 samples with and without indirect call resolution.
Due to page limit, we show the results of 10 representative samples with
real devices to conduct this comparative experiment.
The results are shown in Table~\ref{tab:compOne}.
We illustrate how the indirect call resolution enhances taint analysis and further improves vulnerability discovery from four metrics: the number of covered basic blocks, tainted basic blocks, tainted sinks, and generated alerts.
As clearly shown by the results depicted in Table~\ref{tab:compOne}, 
the number of covered basic blocks increased from 356,585 to 412,688,
and the number of tainted basic blocks increased from 19,218 to 56,488.
This means that indirect call resolution enables \sys to propagate the tainted data to functions where they were not able to reach previously
and taints more variables which were missed before subject to indirect calls. 
The number of tainted sinks increased from 1,234 to 3,710,
which proves that indirect call resolution makes it possible for tainted data to arrive 
unsafe sinks that were unreachable before.
What are the \textbf{benefits} of applying indirect call resolution from the perspective of vulnerability discovery?
As shown in the table, 763 more alerts were produced after resolving functions that were invoked indirectly. 
These alerts in turn correspond to 131 real bugs we introduced before.


For different binaries, 
indirect call resolution has varied impacts on vulnerability detection.
In most cases, indirect call resolution helps report more alerts. 
For Cisco, NERTEAR, Tenda, and TP-Link in our dataset, almost all alerts were discovered with the help of indirect call resolution.
For example,
our prototype produced 0 alert originally without indirect call resolution
and 120 alerts in total afterwards for product R8000 and AC9V3.0.
There is no change in the amount of alerts in firmware sample TEW827RU,
which can be attributed to the fact that
the propagation of tainted data from source to sink does not involve any indirect calls
in this sample.

To further evaluate the effectiveness of indirect call resolution,
we manually analyzed the 162 real bugs (11 n-day and 151 0-day) in the 10 samples,
Among them, the trigger path of 131 bugs involves indirect calls.
Although we do not have the ground truth of indirect call targets
in any of these samples (indirect call resolution is widely believed to be difficult,
even if the source code is available) and thus cannot
evaluate the accuracy of the recovered targets,
these 131 real bugs demonstrate the importance of indirect call resolution.

\begin{table*}[t!]
    \centering
    \setlength{\abovecaptionskip}{1.0mm}
    \caption{Comparison with \textit{KARONTE} and \textit{SaTC} on the same dataset. }
    \begin{adjustbox}{max width=0.99\textwidth}
    \begin{tabular}{l r r r r r r r r r r r r r}
        \toprule

        \multirow{2}{*}{\textbf{Vendor}} & \multirow{2}{*}{\textbf{Samples}}  & 
        \multicolumn{4}{c}{\textbf{\textit{KARONTE}}~\cite{redini2020karonte}} & 
        \multicolumn{4}{c}{\textbf{\textit{SaTC}}~\cite{chen2021sharing}} & 
        \multicolumn{4}{c}{\textbf{\sys}} \\

        & & \textbf{Alerts} & \textbf{\# of KTP} & \textbf{KTP Rate} & \textbf{Time} & \textbf{Alerts} & \textbf{\# of KTP} & \textbf{KTP Rate} & \textbf{Time} & \textbf{Alerts} & \textbf{\# of KTP} & \textbf{KTP Rate} & \textbf{Time} \\

        \midrule 

        NETGEAR & 17 & 36 & 23 & 63.9\% & 17:13 & 1,901 & 537 & 28.2\% & 16:47 & 849 & 849 & 100.0\% & 00:05 \\

        D-Link & 9 & 24 & 15 & 62.5\% & 14:09 & 32 & 22 & 68.8\% & 01:57 & 299 & 234 & 78.3\% & 00:02 \\

        TP-Link & 16 & 2 & 2 & 100.0\% & 01:30 & 7 & 2 & 28.6\% & 04:13 & 73 & 73 & 100.0\% & 00:05 \\

        Tenda & 7 & 12 & 6 & 50.0\% & 01:01 & 144 & 122 & 84.7\% & 12:19 & 362 & 362 & 100.0\% & 00:05 \\

        \hline

        Total & 49 & 74 & 46 & 62.2\% & 451:06 & 2,084 & 683 & 32.8\% & 459:33 & 1,583 & 1,518 & 95.9\% & 03:38 \\

        \bottomrule 
    \end{tabular}
    \label{tab:compTwo}
    \end{adjustbox}
\flushleft
\scriptsize{
For each tool, we report
the total number of generated alerts, the number of true positives based
on the definition in \textit{KARONTE} (\# of KTP), the true-positive rate (KTP Rate),
and the average analysis time for each sample (hh:mm). In the row labelled ``Total'',
we show the aggregated time to analyze all 49 samples in the column ``Time''.
}
\end{table*}

\vspace{-5mm}
\subsection{Comparison with KARONTE and SaTC}
\label{section:comp2}
\vspace{-2mm}


Three works are closely related to ours, including CodeSonar~\cite{codesonar}, 
KARONTE~\cite{redini2020karonte} and SaTC~\cite{chen2021sharing}.
However,
CodeSonar is a commercial product of GrammaTech
and we were unable to use it in our evaluation.
KARONTE and SaTC are both open-sourced.
They perform taint analysis to detect the taint-style vulnerability in 
embedded firmware through the under-constrained symbolic execution on top of
angr~\cite{shoshitaishvili2016sok}.
Due to the very similar scope, we reused 
the dataset~\cite{Kdataset} released by KARONTE,
which includes 49 firmware samples from 4 embedded vendors
(\ie~NETGEAR, D-Link, TP-Link and Tenda).
In fact, SaTC also used this dataset.
This made our experiment a head-to-head comparison.


Table~\ref{tab:compTwo} shows the results,
where the information for KARONTE and SaTC are copied directly from the
original papers~\cite{redini2020karonte,chen2021sharing}.
To be fair, we adopt the same definition of true positive in KARONTE
(\textit{cf.} $\S$X.D in \cite{redini2020karonte}).
Specially, an alert is a true positive if the tainted data that reaches the 
sink is provided by the user. This applies to SaTC too.
KARONTE took approximately 451 hours to produce 74 alerts, among which 46 were true positives;
SaTC took approximately 459 hours to produce 2,084 alerts, among which 683 were true positives;
\sys reported 1,583 alerts in less than 4 hours,
1,518 of which were true positives.
The result shows that \sys can find more true positives in less time than KARONTE and SaTC.
Because of the improvement over existing work,
our tool found 22 new 0-day vulnerabilities in the already extensively
tested samples. They are included
as part of the 151 0-day vulnerabilities we mentioned before.

Apart from the lack of indirect call resolution and less accurate alias
analysis, we found that KARONTE and SaTC frequently raise false alerts
because of the incorrectly specified taint sources. Specifically, due to the
path explosion problem of symbolic execution, they cannot directly use the
commonly recognized taint sources such as \texttt{recv}. Instead, to shorten
the path from the sources to sinks, KARONTE
infers the taint source through the interaction between the back-end programs
in the firmware with a preset list of network-encoding strings
(\eg~``soap'' or ``HTTP''), and SaTC infers the taint sources by the keywords
shared between the front-end files and the back-end programs. When these
keywords are unrelated to user inputs, false positives occur. In
contrast, \sys starts taint analysis directly from the commonly used sources
summarized in Table~\ref{tab:source}, which helps us achieve much higher true positive
rates. We discuss false positives introduced 
by \sys in Section~\ref{sec:limitaiton}.

\section{Limitations}
\label{sec:limitaiton}

In this section, we discuss the limitations of the proposed approach.
SSE, the fundamental technique of the proposed system,
works well to track pointer between indirect memory accesses.
However, it falls short when the pointer involves bitwise operations or 
the offset of the indirect memory access is not a constant.
As such, false negatives could occur for aliases introduced by
these operations.
Further study is needed to improve SSE to handle this situation.


Second, the current prototype supports discovering buffer overflow and command
injection vulnerabilities. In our experiment, most discovered buffer overflow
bugs were stack based. Our tool stayed silent in the presence of many heap
overflow bugs (false negative). To more accurately predict heap overflows,
precise analysis is needed to determine the size of the dynamically allocated
heap buffer.

Lastly, our tool generates false positives (14\% based on our evaluation).
Although false positive is fundamental to all static approaches, we explain
the root causes specific to our method and discuss how to improve on it.
First, our tool misses some constraint checks already exists in the firmware.
This is because
our prototype analyzes library functions by manually generating function summaries.
This leads to missing security checks that do occur in library functions without summaries.
Generating more summaries can mitigate this limitation.
Second, our tool may incorrectly recover indirect calls that do not exist.
This leads to infeasible paths.
Using dynamic analysis to filter the results 
(\eg~dynamic symbolic execution can be used to solve the path constraints) might help reduce 
this kind of false positives.
We leave this as our future work.
Third, \sys uses the \texttt{read} function as a taint source.
However, data read from local files actually cannot be manipulated by attackers.
Although \sys filters obvious read operations from local files,
there are cases that \sys cannot distinguish statically.



\section{Related Work}


\paragraph{Vulnerability discovery using static analysis.}
In static binary analysis, 
most work detects vulnerabilities through data flow analysis.
Balakrishnan et al. first proposed
using VSA to achieve 
flow-, context-sensitive data flow analysis~\cite{balakrishnan2010wysinwyx}
and further commercialized the idea to implement a product
called CodeSurfer/x86~\cite{balakrishnan2005codesurfer}.
Similar work has been proposed to find taint-style vulnerabilities
using VSA~\cite{rawat2011static, rawat2014listt,cheng2011loongchecker}.
However, VSA cannot achieve high field sensitivity in the multi-level pointers 
when the address of memory access cannot be concretized.
Some work~\cite{cova2006static, redini2017bootstomp, redini2020karonte,chen2021sharing} 
was proposed to discover taint-style vulnerabilities through tracking data-flow information
with symbolic execution.
Among them, KARONTE~\cite{redini2020karonte} is a static analysis framework
for embedded firmware that can discover vulnerabilities
due to multi-binary interactions. The authors achieve
this goal by modeling and tracking multi-binary interactions.
SaTC~\cite{chen2021sharing} also performs taint analysis to discover bugs. It utilizes shared keywords related to user input in the front-end and back-end to infer the taint source.
Ivan Gotovchits et al., proposed an approach for collecting path- and context-sensitive
data-flow information called $\mu$\textit{flux}.  
$\mu$\textit{flux} can find taint-style vulnerabilities by static property checking~\cite{gotovchits2018saluki}. 
In order to find more taint-style vulnerabilities,
DTaint~\cite{cheng2018dtaint} adopts
pointer alias analysis to improve the data flow analysis and 
utilizes data structure similarity matching
to construct data dependence between functions invoked by indirect calls.
However, DTaint lacks accuracy and efficiency in data flow analysis.
Most of the static approaches for binary
cannot maintain field-sensitivity for data-flow analysis.
All the aforementioned works that detect taint-style vulnerabilities
did not perform indirect call resolution, which we have demonstrated
to be critical. \sys addresses both problems.

\paragraph{Alias analysis.}
Alias analysis is a long-term research topic in static binary analysis.
Debray et al.~\cite{debray1998alias} proposed an inter-procedural flow-sensitive pointer alias analysis for \textit{x}86 executables, which is context-\textbf{in}sensitive.
Guo et al.~\cite{guo2005practical} presented the first context-sensitive points-to analysis for \textit{x}86 assembly code.
However, this approach is only partially flow-sensitive.
Reps et al.~\cite{reps2008improved} utilized  value-set analysis (VSA) to identify pointer alias through tracking memory accesses in \textit{x}86 executables.
Chong et al.~\cite{chong2013accurate} presented a flow-sensitive algorithm for instruction-level alias analysis on ARM executables, which is context-\textbf{in}sensitive.
We proposed a new alias analysis technique based on structured symbolic expressions (SSE).
Our implementation is based on VEX IR.
Therefore it is architecture-neutral and we have applied our tools
against both ARM and MIPS executables.
To the best of our knowledge, this is the first work
that simultaneously achieves demand-driven, flow-, context- and field-sensitive 
alias analysis for binary.
SSE shares the spirit of access path~\cite{cheng2000modular}. 
However, access path targets source code, not the binary.






\paragraph{Vulnerability discovery  using dynamic analysis.} 
Fuzzing has been widely adopted as a dynamic analysis technique
to uncover vulnerabilities in embedded devices. 
RPFuzzer~\cite{wang2013rpfuzzer} is a
fuzzing framework specifically designed for finding protocol vulnerabilities
for router devices.
IoTFuzzer~\cite{chen2018iotfuzzer} leverages
the companion mobile apps of IoT devices to perform efficient black-box fuzzing.
FIRM-AFL~\cite{zheng2019firm} was presented as a high-throughput greybox fuzzer for firmware running a POSIX-compatible operating system through augmented process emulation.
These emulation-based fuzzing approaches rely on correctly executing
the firmware in an emulator in the first place.
However, accurate emulation is a non-trivial task in practice due to the
diversity of embedded devices~\cite{muench2018you}.

\vspace{-4mm}
\section{Conclusion}

In this work, we propose a precise demand-driven flow-, context- and field-sensitive alias analysis approach.
Based on this new approach, we also design a novel indirect call resolution scheme.
Combined with sanitization rule checking, our solution 
discovers taint-style vulnerabilities by static taint analysis.
We implemented our idea with a prototype called \sys and
evaluated it against 35 real-world embedded firmware samples from six popular vendors.
\sys discovered at least 192 bugs, including 41 n-day bugs and 151 0-day bugs.
At least 115 CVE/PSV numbers have been allocated from a subset of the reported vulnerabilities at the time of writing.
Compared to the state-of-the-art tools such as KARONTE and SaTC,
\sys found significantly more bugs on the same dataset in less time.

{
\small
\bibliographystyle{plain}
\begin{spacing}{0.9}
\bibliography{References}
\end{spacing}
}

\appendix

\begin{table}[!hptb]
    \centering
    \setlength{\abovecaptionskip}{1.0mm}
    \caption{Taint sources and sinks.}
    \begin{adjustbox}{max width=0.9\columnwidth}
    \begin{tabular}{cc}
        \toprule
        \multirow{2}{*}{\textbf{Taint Sources}} & recv, recvfrom, read, fread, fgets \\
         & BIO\_read, BIO\_gets, SSL\_read, getenv \\
         \hline
         \multirow{3}{*}{\textbf{Taint Sinks}} & strcpy, strncpy, memcpy, memmove \\
          & sprintf, sscanf, strcat, strncat \\
          & system, popen, execve \\
        \bottomrule 
    \end{tabular}
    \label{tab:source}
    \end{adjustbox}
    \vspace{-5mm}
\end{table}

\begin{table}[!hptb]
    \centering
    \setlength{\abovecaptionskip}{1.0mm}
    \caption{The summaries of library functions that propagate tainted data.}
    \begin{adjustbox}{max width=0.9\columnwidth}
    \begin{tabular}{c c}
        \toprule
        \multirow{3}{*}{\textbf{String Copy}} & strcpy, strncpy, strlcpy, memcpy,  \\
        & memmove, sprintf, snprintf, vsnprintf \\
        & strcat, strncat, sscanf, strdup \\
        \textbf{String Index} & strstr, strchr, strrchr, strpbrk, stristr \\
        \textbf{String Split} & strtok, strtok\_r, strsep \\
        \textbf{String to Int} & atoi, atol, atoll, strtol, strtoll, strtoul \\
        \textbf{Other functions} & hsearch\_r, index, strlen \\
        \bottomrule 
    \end{tabular}
    \label{tab:lib-tran}
    \end{adjustbox}
\end{table}

\begin{table}[t]
    \centering
    \setlength{\abovecaptionskip}{1.0mm}
    \caption{\label{tab:OneDay}N-day vulnerabilities} 
    \begin{adjustbox}{max width=0.8\columnwidth}
    \begin{tabular}{lcc}
        \toprule
        \textbf{Vendor} & \textbf{Vulnerability IDs} & \textbf{Alerts}  \\
        \midrule 
        \textbf{Cisco} & CVE-2019-1663 CVE-2019-1652 & 2 \\
        \hline
        \multirow{4}{*}{\textbf{TRENDnet}} & CVE-2019-13151 CVE-2019-13278 & \multirow{4}{*}{68} \\
        & CVE-2019-13276 CVE-2019-13279 & \\
        & CVE-2019-13280 CVE-2019-13150 & \\
        & CVE-2019-11418 & \\
        \hline
        \multirow{4}{*}{\textbf{D-Link}} & CVE-2015-2052 CVE-2013-7389 & \multirow{4}{*}{9} \\ 
        & CVE-2016-5681 CVE-2015-2051 & \\
        & CVE-2018-17787 CVE-2019-9122 & \\
        & CVE-2018-6530 CVE-2019-10999 &  \\
        & CVE-2019-19597 & \\
        \hline
        \multirow{2}{*}{\textbf{Netgear}} & CVE-2017-6077 CVE-2017-6334 &  \multirow{2}{*}{3} \\
        & EDB-ID:43055  &  \\
        \hline
        \multirow{9}{*}{\textbf{Tenda}} & CVE-2018-18732 CVE-2020-13390 &  \multirow{9}{*}{36} \\
        & CVE-2020-13391 CVE-2020-13392 & \\
        & CVE-2020-13393 CVE-2020-13394 & \\
        & CVE-2018-14559 CVE-2018-14492 & \\
        & CVE-2018-14557 CVE-2018-16333 & \\
        & CVE-2018-18706 CVE-2018-18707 & \\
        & CVE-2018-18708 CVE-2018-18709 & \\
        & CVE-2018-18727 CVE-2018-18730 & \\
        & CVE-2018-18731 CVE-2020-13389 & \\
        \hline
        \multirow{1}{*}{\textbf{TP-Link}} & CVE-2018-16119 CVE-2017-13772 &  \multirow{1}{*}{2} \\
        \hline
        \textbf{Total} & \textbf{41} & \textbf{120} \\
        \bottomrule 
    \end{tabular} 
    \end{adjustbox}
    \vspace{-1mm}
\end{table}

\begin{table*}[h!]
    \centering
    \setlength{\abovecaptionskip}{1.0mm}
    \caption{\label{tab:0day}Zero-day vulnerabilities found by \sys} 
    \begin{adjustbox}{max width=0.98\textwidth}
    \begin{tabular}{lcc}
        \toprule
        \textbf{Vendor} & \textbf{Model} & \textbf{Vulnerability IDs}  \\
        \midrule 

        \multirow{14}{*}{Cisco} 
        & \multirow{8}{*}{RV320}
        & CVE-2020-3274, CVE-2020-3275, CVE-2020-3276, CVE-2020-3277, CVE-2020-3278, CVE-2020-3279, CVE-2020-3286 \\
        &
        & CVE-2020-3287, CVE-2020-3288, CVE-2020-3289,CVE-2020-3290, CVE-2020-3291, CVE-2020-3292, CVE-2020-3293 \\
        & 
        & CVE-2020-3294, CVE-2020-3295, CVE-2020-3296, CVE-2021-1319, CVE-2021-1320, CVE-2021-1321,CVE-2021-1322 \\
        &
        & CVE-2021-1323, CVE-2021-1324, CVE-2021-1325, CVE-2021-1326, CVE-2021-1327, CVE-2021-1328, CVE-2021-1329 \\
        &
        & CVE-2021-1330, CVE-2021-1331, CVE-2021-1332, CVE-2021-1333, CVE-2021-1334, CVE-2021-1335, CVE-2021-1336 \\
        &
        & CVE-2021-1337, CVE-2021-1338, CVE-2021-1339, CVE-2021-1340, CVE-2021-1341,CVE-2021-1342, CVE-2021-1343 \\
        &
        & CVE-2021-1344, CVE-2021-1345, CVE-2021-1346, CVE-2021-1347, CVE-2021-1348, CVE-2021-1314, CVE-2021-1315 \\
        &
        & CVE-2021-1316, CVE-2021-1317, CVE-2021-1318 \\
        \cline{2-3}
        & \multirow{6}{*}{RV130} 
        & CVE-2020-3268, CVE-2020-3269, CVE-2021-1146, CVE-2021-1147, CVE-2021-1148, CVE-2021-1150, CVE-2021-1159 \\
        &
        & CVE-2021-1160, CVE-2021-1161, CVE-2021-1162, CVE-2021-1163, CVE-2021-1165, CVE-2021-1166, CVE-2021-1169 \\
        &
        & CVE-2021-1170, CVE-2021-1171, CVE-2021-1172, CVE-2021-1173, CVE-2021-1174, CVE-2021-1175, CVE-2021-1176 \\
        &
        & CVE-2021-1177, CVE-2021-1178, CVE-2021-1179, CVE-2021-1180, CVE-2021-1181, CVE-2021-1182, CVE-2021-1183 \\
        &
        & CVE-2021-1184, CVE-2021-1185, CVE-2021-1186, CVE-2021-1187, CVE-2021-1188, CVE-2021-1189, CVE-2021-1190 \\
        &
        & CVE-2021-1191, CVE-2021-1192, CVE-2021-1193, CVE-2021-1194, CVE-2021-1195, CVE-2021-1196, CVE-2021-1203, CVE-2021-1204 \\

        \hline

        \multirow{2}{*}{D-Link} 
        & DIR-825
        & CVE-2020-10213, CVE-2020-10215, CVE-2020-10214, CVE-2020-10216 \\
        \cline{2-3}
        & DAP-1860
        & 2 unassigned \\

        \hline

        \multirow{3}{*}{TRENDnet}
        & TEW632BRP
        & CVE-2020-10213, CVE-2020-10215, CVE-2020-10216 \\
        \cline{2-3}
        & \multirow{2}{*}{TEW827DRU}
        & CVE-2020-14074, CVE-2020-14075, CVE-2020-14076, CVE-2020-14077, CVE-2020-14078, CVE-2020-14079, CVE-2020-14080, CVE-2020-14081 \\
        &
        & 14 unassigned \\

        \hline

        \multirow{3}{*}{NETGEAR}
        & R7800 
        & 1 unassigned \\
        \cline{2-3}
        & \multirow{2}{*}{R8000}
        & PSV-2020-0300,  PSV-2020-0315,  PSV-2020-0314, PSV-2020-0312, PSV-2020-0311, PSV-2020-0310, PSV-2020-0309, PSV-2020-0308 \\
        & 
        & 13 unassigned \\

        \hline

        TP-Link & WR940NV4 & 2 unassigned \\

        \hline

        Tenda & AC9V3 & 4 unassigned \\

        \hline

        \textbf{Total} & \textbf{10} & \textbf{151 vulnerabilities, 115 assigned with public exposure IDs} \\
        \bottomrule 
    \end{tabular} 
    \end{adjustbox}
\vspace{-4mm}
\end{table*}


\section{Taint Source and Taint Sink}
\label{source-sink}

Table~\ref{tab:source} shows all sources and sinks used in \sys.


\section{Library Function Summaries}
\label{lib-summar}

Table~\ref{tab:lib-tran} shows the summaries of 29 common functions from the Standard C Library.


\section{Vulnerability Discovery}
\label{Ap-alerts}

Table~\ref{tab:vul2} shows the alerts produced by \sys for the remaining 25 of 35 firmware samples.
The results of the other 10 samples are shown in Table~\ref{tab:vul} in Section~\ref{sec:eval:vul}. 
Table~\ref{tab:OneDay} shows the n-day vulnerabilities found by \sys for 35 firmware samples.
Table~\ref{tab:0day} shows the 0-day vulnerabilities found by \sys for 35 firmware samples.

\section{Indirect Call Resolution}
\label{Ap-icalls}

Table~\ref{tab:icall2} shows the results of indirect call resolution for the remaining 25 of 35 firmware samples.
The results of the other 10 samples are shown in Table~\ref{tab:icall} in Section~\ref{section:icall}.

\begin{table*}[t]
    \centering
    \setlength{\abovecaptionskip}{1.0mm}
    \caption{\label{tab:vul2} Alerts produced by \sys for the remaning 25 firmware samples} 
    \begin{adjustbox}{max width=0.98\textwidth}
    \begin{tabular}{lccccrrrrr} 
        \toprule    
        \textbf{Vendor} & \textbf{ID} & \textbf{Firmware Version} & \textbf{Arch} & 
        \textbf{Binary} & \textbf{Size (KB)} & \textbf{Ana. Func} & \textbf{Tainted Sinks} & \textbf{Alerts} & \textbf{Time (s)} \\ 
        \midrule 

        \multirow{5}{*}{\textbf{D-Link (5)}}
        & 5 & DIR-645\_1.03 & MIPS3 & cgibin & 156 & 190  & 37 & 21 & 18.85 \\
        & 6 & DIR-890L\_A1\_1.03 & ARM32 & cgibin & 151  & 305  & 49 & 21 & 42.26 \\ 
        & 7 & DIR-868L\_A1\_b04 & ARM32 & cgibin & 151 & 368  & 43 & 8 & 38.63 \\
        & 8 & DIR-823G\_A1\_B03 & MIPS32  & goahead & 1,525 & 1043  & 52 & 5 & 82.18  \\
        & 9 & DCS-5020L\_A1\_v1.15.12 & MIPS32 & alphapd & 707 & 1,086  & 48 & 4  & 716.65 \\ \addlinespace
        
        
        \multirow{15}{*}{\textbf{NETGEAR (15)}}
        & 14 & R6200v2\_v1.0.3.12 & ARM32 & httpd & 1,259 & 852 & 257 & 52 & 123.00  \\
        & 15 & R6300v2\_v1.0.4.18 & ARM32 & httpd & 1,294 &  864  & 254 & 24 & 132.70  \\
        & 16 & R6400\_v1.0.1.46 & ARM32 & httpd & 1,482 & 1,123 &  401 & 36 &  148.74 \\
        & 17 & R6700\_v1.0.1.36 & ARM32 & httpd & 1,762 & 1,320  & 464 & 36  & 273.97  \\
        & 18 & R7000P\_v1.3.0.8 & ARM32 & httpd & 1,764 & 1,349 & 480 & 39  & 288.00 \\ 
        & 19 & R7300DST\_v1.0.0.56 & ARM32 & httpd & 1,452 & 1,030 & 386 & 35 & 184.38  \\
        & 20 & R7500v2\_v1.0.3.16 & ARM32 & net-cgi & 603 & 1,473  & 226 & 75  & 110.89 \\
        & 21 & R7900\_v1.0.1.26 & ARM32 & httpd & 1,499 & 1,080   & 425 & 37  & 209.03 \\
        & 22 & R8300\_v1.0.2.106 & ARM32 & httpd & 1,490 & 1,043  & 393 & 37 & 152.15  \\
        & 23 & R9000\_v1.0.2.40 & ARM32 & net-cgi & 635 & 1,294  & 227 & 94 & 89.40  \\
        & 24 & DGN1000\_v1.1.00.46 & MIPS32 & setup.cgi & 324 &  732 & 88 & 56 & 30.55   \\
        & 25 & DGN2200\_v1.0.0.50 & MIPS32 & httpd & 990 & 773  & 257 & 69  & 111.27 \\ 
        & 26 & AC1450\_v1.0.0.36 & ARM32 & httpd & 1,230 & 835 & 246 & 49  & 92.70  \\
        & 27 & WNR3500Lv2\_v1.2.0.46 & MIPS32 & httpd & 1,584 & 888  & 375 & 14 & 202.17  \\
        & 28 & XR500\_v2.1.0.4 & ARM32 & net-cgi & 536 & 1,327  & 152 & 35 & 81.70  \\ \addlinespace
        
        \multirow{2}{*}{\textbf{TP-Link (2)}}
        & 30 & MR3020NV1\_en.3.17.2 & MIPS32 & httpd & 1,524 & 2,613  & 264 & 10 & 208.28 \\
        & 31 & WR1043NDV3\_en.3.16.9 & MIPS32 & httpd & 1,900 & 3,629 & 254 & 32 & 373.21  \\ \addlinespace
        
        \multirow{3}{*}{\textbf{Tenda (3)}}
        & 33 & AC10V1.0RTL\_v15.03.06.23 & MIPS32 & httpd & 2,011 & 1,187  & 166 & 82 & 358.26 \\
        & 34 & AC15V1.0BR\_v15.03.05.18 & ARM32 & app\_data\_center & 83 & 182  & 16 & 5 & 10.94 \\
        & 35 & WH450AV1BR\_v1.0.0.18 & MIPS32 & httpd & 416 & 564  & 76 & 40 & 194.42  \\
        
        \bottomrule 
    \end{tabular} 
    \end{adjustbox}
    \vspace{-5.0mm}
\end{table*}

\begin{table*}[t] 
    \centering
    \setlength{\abovecaptionskip}{1.0mm}
    \caption{\label{tab:icall2}Results of indirect call resolution for the remaning 25 firmware samples} 
    \begin{adjustbox}{max width=0.65\textwidth}
    \begin{tabular}{llrrrrr} 
        \toprule   

        \multirow{2}{*}{\textbf{ID}} & \multirow{2}{*}{\textbf{Model}} & \textbf{All} & 
        \textbf{Resolved} & \textbf{I-Call} &  \textbf{\% of resolved } & \textbf{Time} \\

         & & \textbf{I-Calls} & \textbf{I-Calls} & \textbf{targets} & \textbf{I-Calls} & \textbf{(sec.)} \\

        \midrule 

         5 & D-Link DIR-645  & 10 & 7  & 47 &  70.0\% & 4.82  \\
         6 & D-Link DIR-890L & 5 & 4 & 24 &  80.0\% & 3.60  \\ 
         7 & D-Link DIR-868L & 5 & 4 & 23  & 80.0\% & 3.70  \\
         8 & D-Link DIR-823G & 43 & 25 & 399  & 58.1\% & 54.25  \\
         9 & D-Link DCS-5020L & 94 & 73 & 478 &  76.6\% & 410.10  \\ \addlinespace%
        
         14 & NETGEAR R6200v2 & 3 & 2  & 410  &  66.6\% & 27.20  \\
         15 & NETGEAR R6300v2  & 3 & 2  & 415 &  66.6\% & 28.67  \\
         16 & NETGEAR R6400 & 3 & 2  & 486  &  66.6\% & 38.06  \\
         17 & NETGEAR R6700 & 3 & 2  & 612 &  66.6\% & 66.83  \\
         18 & NETGEAR R7000P & 3 & 2  & 617 &  66.6\% & 73.21  \\
         19 & NETGEAR R7300DST & 3 & 2  & 492  &  66.6\% & 37.44  \\
         20 & NETGEAR R7500v2 & 28 & 25  & 807 &  89.2\% & 40.21  \\
         21 & NETGEAR R7900  & 3 & 2  & 484 &  66.6\% & 37.72  \\
         22 & NETGEAR R8300 & 3 & 2  & 496 & 66.6\% & 38.36  \\
         23 & NETGEAR R8900 & 52 & 47  & 581 &  90.3\% & 36.87  \\
         24 & NETGEAR DGN1000 & 40 & 36  & 642 &  90.0\% & 26.33  \\ 
         25 & NETGEAR DGN2200 & 3,206 & 3,204  & 476 &  99.9\% & 130.91  \\  
         26 & NETGEAR AC1450 & 3 & 2 & 414 &  66.6\% & 30.46  \\
         27 & NETGEAR WNR3500Lv2 & 4,667 & 4,662  & 574 & 99.8\% & 161.06 \\ 
         28 & NETGEAR XR500 & 29 & 13  & 629 &  44.8\% & 32.27  \\ \addlinespace%
        
         30 & TP-Link MR3020NV1 & 299 & 236  & 553 &  78.9\% & 170.53  \\
         31 & TP-Link WR1043NDV3 & 388 & 308  & 687 &  79.4\% & 431.28  \\  \addlinespace%
        
         33 & Tenda AC10V1.0RTL & 88 & 65  & 279 &  73.9\% & 208.19  \\
         34 & Tenda AC15V1.0BR & 47 & 38  & 2 & 80.8\% & 4.51 \\
         35 & Tenda WH450AV1BR & 2,063 & 2,043  & 269 &  99.0\% & 118.16 \\
        \bottomrule 
    \end{tabular} 
    \end{adjustbox}
    \vspace{-5.0mm}
\end{table*}

\end{document}